\documentclass[fleqn,usenatbib]{mnras}
\usepackage[utf8]{inputenc}
\usepackage{newtxtext,newtxmath}

\usepackage[T1]{fontenc}
\usepackage{ae,aecompl}
\usepackage{comment}

\usepackage{natbib}

\usepackage{graphicx}	
\usepackage{newtxtext,newtxmath}

\usepackage{xspace}
\newcommand{\ltw}{low-$T/|W|$\xspace}
\newcommand{\BV}{Brunt-V\"ais\"al\"a }

\usepackage{bm}
\renewcommand{\vec}[1]{\bm{#1}}
\newcommand{\tensor}[1]{\bm{#1}}

\newcommand{\taki}[1]{\textcolor{red}{#1}}

\title[Insights into 3D rotating supernova models]
{Insights into non-axisymmetric instabilities
 in three-dimensional rotating supernova models with neutrino and gravitational-wave signatures}

\author[T. Takiwaki, K.Kotake, and T.Foglizzo]{
Tomoya Takiwaki$^{1}$,
Kei Kotake$^{2,3}$,
and Thierry Foglizzo$^{4}$
\\$^{1}${Division of Science, National Astronomical Observatory of Japan, 2-21-1, Mitaka, Tokyo 181-8588, Japan}
\\$^{2}${Department of Applied Physics, Fukuoka University, 8-19-1, Nanakuma, Johnan, Fukuoka 814-0180, Japan}
\\$^{3}${Research Institute of Stellar Explosive Phenomena, Fukuoka University, 8-19-1, Nanakuma, Jonan, Fukuoka 814-0180, Japan}
\\$^{4}${Laboratoire AIM (CEA/Irfu, CNRS/INSU, Univ. Paris Diderot), CEA Saclay, F-91191 Gif sur Yvette, Cedex, France}
}

\date{Accepted XXX. Received YYY; in original form ZZZ}

\begin{document}
\label{firstpage}
\pagerange{\pageref{firstpage}--\pageref{lastpage}}
\maketitle

\begin{abstract}
We present a detailed analysis to clarify what determines the growth of the low-$T/|W|$ instability in the context of rapidly rotating core-collapse of massive stars. To this end, we perform three-dimensional core-collapse supernova (CCSN) simulations of a $27 M_{\odot}$ star including several updates in the general relativistic correction to gravity, the multi-energy treatment of heavy-lepton neutrinos, and the nuclear equation of state. 
Non-axisymmetric deformations are analyzed from the point of view of the time evolution of the pattern frequency and the corotation radius. The corotation radius is found to coincide with the convective layer in the proto neutron star (PNS).
We propose a new mechanism to account for the growth of the low-$T/|W|$ instability in the CCSN environment.
Near the convective boundary where a small Brunt-V\"ais\"al\"a frequency is expected,
Rossby waves propagating in the azimuthal direction at mid latitude induce non-axisymmetric unstable modes, in both hemispheres.
They merge with each other and finally become the spiral arm in the equatorial plane.
We also investigate how the growth of the low-$T/|W|$ instability impacts the neutrino and gravitational-wave signatures.
\end{abstract}

\begin{keywords}
stars: massive -- supernovae: general -- neutrinos -- gravitational waves
\end{keywords}



\section{Introduction}\label{sec:introduction}

The theoretical understanding of the explosion of massive stars depends on the development of hydrodynamic instabilities in three dimensions (3D).
A growing number of core-collapse supernova (CCSN) simulations have been
 performed in 3D with sophisticated neutrino transport schemes. Some of the models successfully obtained the 
 shock revival owing to the neutrino mechanism, leading to the onset of the explosion
 \citep{Takiwaki12,Takiwaki14,Takiwaki16,Melson15a,BMuller15a,lentz15,roberts16,BMuller18,Nakamura19,radice19,Vartanyan19a,Burrows19,Glas19a,Nagakura20,Bollig20}.
 
 These studies suggest that neutrino-driven convection plays a key role to revive the shock.
Some models with high mass accretion rates showed the emergence of the standing accretion shock instability (SASI) \citep{Blondin03,Foglizzo07}.
In 3D simulations of rapidly rotating progenitors, \citet{Summa18} found a vigorous spiral SASI activity, while \citet{Takiwaki16} observed a spiral flow due to the low-$T/|W|$ instability \citep[e.g.][]{Ott05}.
Both obtained successful explosions for progenitors that did not explode in the absence of rotation. 

Gravitational waves (GWs) and neutrinos emitted from the SN core carry the imprints of the dominant multi-dimensional dynamics inside the proto-neutron star (PNS, e.g. \cite{ernazar2020,horiuchi18} for recent review).
Measuring these signals would provide observational constraints on the inner-workings of the supernova engine.
Some of the previous studies reported that the SASI-dominated models showed a time variability with relatively low frequency ($< 200$\,Hz) in both GWs \citep{andresen19,Mezzacappa20} and neutrinos \citep{Walk18,Walk20,Nagakura20}. The SASI-induced neutrino modulation is known to be highly dependent on the observer's direction  \citep[e.g.][]{Tamborra14prd}.
There were, however, no clear low-frequency and angle-dependent feature in GWs and neutrinos for the convection-dominated models \citep{KurodaT18,Vartanyan19b,powell19,Powell20}. Regardless of the presence or the absence of the SASI, it is interesting that the lepton emission self-sustained asymmetry (LESA) first found in \citet{Tamborra14ApJ} also potentially leads to anisotropic emission of electron-type and anti-electron type neutrinos \citep{O'Connor18,Glas19b,Walk19,Vartanyan19b,Walk20}. 

Rapid rotation of the precollapse core also leads to angle-dependent features of GW and neutrino emission.
The rapidly rotating model in \citet{Takiwaki18} that explodes due to the low-$T/|W|$ instability exhibited a clear quasi-periodic modulation of the neutrinos  if the observer is on an equatorial plane perpendicular to the rotation axis. Toward the polar direction, the GW emission is strongest with the GW frequency about twice higher than the (rotation-induced) neutrino modulation frequency (see also \citet{Shibagaki20} for the black-hole forming case).

The \ltw instability in the supernova core (i.e. in the accreting PNS) 
is a relatively new subject, and little is known about its mechanism. Most of our knowledge is based on stability analyses in isolated rotating stars, assuming a polytropic equation of state \citep{Centrella01,Shibata03a,Saijo03,Karino03,Watts05,Ou06,Saijo06,Passamonti15,passa20}.
The \ltw instability was named after its lower threshold value of the ratio of the rotation energy $T$ to the gravitational energy $W$, 
compared to the onset $T/|W|\sim 0.27$ of the dynamical bar mode instability.
A pioneering work of \cite{Centrella01} obtained the threshold value of $T/|W|\sim 0.14$ and 
\cite{Shibata03a} extensively performed 3D simulations and found that threshold $T/|W|$ can be $\sim 0.01$ with strong differential rotation.
\cite{Yoshida17} have proposed a mechanism of the instability based on the over-reflection of sound waves, where non-axisymmetric sound waves are trapped
between the surface of the star and a corotation radius.
In the context of an accreting PNS, a similar instability was found with a ratio $T/|W|$ of a few \% \citep{Ott05,Scheidegger08,Ott12a,Takiwaki16,Shibagaki20}. 
However,
the mechanism may differ from an isolated rotating star since the PNS does not have a sharp surface and is subject to continuous accretion of mass and angular momentum.

In our previous work \citep{Takiwaki16,Takiwaki18}, a Newtonian gravity was assumed and the heavy-lepton ($\nu_X$) neutrinos were treated by a leakage scheme in the 3D simulations. In this work, we include a general relativistic correction and evolve the 
 multi-energy transport of $\nu_X$ neutrinos like electron- and anti-electron neutrinos via the isotropic diffusion source approximation (IDSA) scheme \citep{Liebendorfer09}. We use the equation of state (EOS) from \citet{togashi}, which is more consistent with nuclear experiments and the observational constraints on the neutron star mass-radius relation than the LS220 EOS \citep{Lattimer91} employed in our previous work.
 With this update, we present a detailed analysis to obtain deeper insights toward clarifying what determines the growth of the \ltw instability in the context of rapidly rotating core-collapse. 
 
 This paper is organized as follows. In Section \ref{sec:methods}, we summarize numerical methods and initial models for our 3D core-collapse supernova models. Section \ref{sec3} is devoted to the results which include not only a detailed mode analysis of the PNS deformation to disentangle the various types of hydrodynamics instabilities (such as pressure-, buoyancy-, corotation-driven instabilities), but also the impact on the gravitational-wave and neutrino signatures. In Section \ref{sec4}, we summarize our results and discuss the implications and limitations of this work.

\section{Methods}\label{sec:methods}
 All the 3D models in this work are computed by running
our supernova code, {\small 3DnSNe} \citep{Takiwaki16}.
The code was designed for CCSN radiation-hydrodynamics simulations in a 3D spherical coordinate system ($r$, $\theta$, $\phi$). The spectral neutrino transport is solved
by the IDSA scheme \citep{Liebendorfer09}. We have updated the original two-flavor IDSA scheme (i.e. $\nu_e,\bar{\nu}_e$) in several manners,
such that the evolution of the streaming neutrinos is self-consistently
solved \citep{Takiwaki14} and three-flavor neutrino transport is solved
including approximate general relativistic corrections (e.g. \citealt{Kotake18}
for more details). A detailed code comparison has been performed in
\citet{OConnor18} with one-dimensional (1D) geometry.
The {\small 3DnSNe} code has been widely used in the following studies:
\citet{Nakamura19,Nakamura15,Kotake18} showed the dynamics of supernovae;
\citet{Cherry20, Zaizen20, Sasaki20,Sasaki17} discussed neutrino oscillation properties;
\citet{Sotani20a,Sotani16} provided the linear analysis for the gravitational wave signals.

The basic equations evolved by the {\small 3DnSNe} code are:
\begin{align}
\frac{\partial \rho}{\partial t} &+ \nabla \cdot (\rho \vec{v} ) = 0, \label{basic eq1}\\
\frac{\partial (\rho \vec{v} )}{\partial t} 
&+ \nabla \cdot (\rho \vec{v} \vec{v} + P \tensor{I}
) = - \rho \nabla \Phi, \label{basic eq2}\\
\frac{\partial e}{\partial t} 
&+ \nabla \cdot [(e + P) \vec{v} )]
= - \rho \vec{v} \cdot \nabla \Phi + Q_E, \label{basic eq3}\\
\frac{\partial \rho Y_l}{\partial t} &+ \nabla \cdot (\rho Y_l  \vec{v}) = \Gamma_l \;, \label{basic eq4}\\
\frac{\partial \rho Z_m}{\partial t} &+ \nabla \cdot (\rho Z_m  \vec{v}) + \frac{\rho Z_m}{3}\nabla \cdot \mbox{\boldmath $v$}  = Q_m \;, \label{basic eq5}\\
& \Delta \Phi = 4 \pi G \rho \;, \label{basic eq6}
\end{align}
where $\rho$, $\mbox{\boldmath $v$}$, $P$, $e$,
and $\Phi$ are the mass density, the fluid velocity vector, the total (thermal and magnetic) pressure, the total energy
density and the gravitational potential
 ($G$ is the gravitational constant), respectively. $Y_l$ is the lepton
fraction and the subscript $l$ denotes the species of leptons: $l=e,
\nu_e, \bar{\nu}_e, \nu_X$ and $Z_m$ is the specific internal energy of
the trapped neutrinos and $m$ represents the species of neutrinos:
$m=\nu_e,\bar{\nu}_e,\nu_X$. $Q_E$, $Q_{m}$ are the rates of energy change
and $\Gamma_l$ is the rate of change in the number fraction due to the interaction of the neutrinos
with the fluid. $\mbox{\boldmath{$I$}}$ is the unit matrix. 

To evolve equations (\ref{basic eq1})--(\ref{basic eq5}),
 the HLLC scheme \citep{toro94} is implemented as an approximate Riemann solver. In order to retain the total energy including the gravitational binding energy, we use the method of \cite{Muller10} to solve equation \eqref{basic eq3}. As a solution of equation \eqref{basic eq6}, the spherically symmetric gravitational potential is taken in the form of the
phenomenological general relativistic potential of Case A in \cite{Marek06} 
and the multipolar components are added following \citet{Wongwathanarat2010}.
 
Our setup for the microphysics is similar to \cite{OConnor18}.
The adopted neutrino reaction rate is set5a of \cite{Kotake18}, i.e.
the weak magnetism and recoil correction \citep{Horowitz02} as well as 
nucleon–nucleon bremsstrahlung are added to the standard opacity set
of \cite{Bruenn85}. In this run, 20 energy groups that logarithmically
spread from 1 to 300 MeV are employed. We use the equation of state (EOS)
by \cite{togashi}, which is constructed by the variational many-body theory using realistic nuclear forces. Note that the Togashi EOS is consistent with the requirements from both a nuclear experiment (such as the bulk imcompressibility and the symmetry energy at the saturation density), and the mass-radius relation based on the gravitational-wave detection of the neutron-star binary coalescence
 (see \citet{nakazato20} for collective references therein). 

To see the impact of rotation on the post-bounce dynamics in a controlled manner, we change the progenitor core rotation parametrically taking a non-rotating $27.0$ $M_\odot$ progenitor \citep{Woosley02}.
Following our previous studies \citep{Takiwaki16,Takiwaki18}, the constant angular frequency $\Omega_0$ is initially imposed to the iron core with a cut-off ($\propto r^{-2}$) outside. The three models referred to as R0.0, R1.0, and R2.0 correspond to $\Omega_0 = 0,1,2$ rad/s, respectively. The dimension of the space is also indicated, e.g. R2.0-3D.

The grid spacing in this work is similar to the 3D runs in
\citet{Takiwaki18}. In the radial direction, a logarithmically stretched
grid is adopted with $512$ zones from the center up to $5000$ km,
whereas the polar angle in the $\theta$-direction is uniformly divided into
$\Delta \theta = \pi /64$ (for the 2D models $\Delta \theta = \pi /128$).
Similarly the azimuthal angle  in the $\phi$-direction 
 is uniformly divided into $\Delta \phi = 2\pi /128$.
The innermost $10$ km are computed in spherical
symmetry to avoid excessive time-step limitations. Reflective boundary
conditions are imposed at the inner radial boundary ($r=0$), while fixed-boundary
conditions are adopted at the outer radial boundary ($r=5000$ km) except
the gravitational potential that is inversely proportional to the radius
at the outer ghost cells. Time is measured after bounce ($t_{\rm pb} = 0$), until otherwise stated.

\section{Results}\label{sec3}
The most prominent feature in our models is the development of a non-axisymmetric instability which we refer to as the \ltw instability.
In Section \ref{sec:overall}, we demonstrate how the overall dynamics is affected by this instability.
As presented in \cite{Takiwaki16}, this instability not only assists the onset of the shock revival, but also significantly affects the morphology of the shock.
In Section \ref{sec:detailofTW}, we investigate the features of the instability in detail. Comparing the pattern frequency and the hydrodynamic angular frequency, we found the coincidence of the corotation radius and the convective layer and propose a new scenario for the growth of the instability.
Finally in Section \ref{sec:GW_nu}, we address the neutrino- and gravitational-wave emission signatures induced by the instability.
This last section is an extension of \cite{Takiwaki18}.

\subsection{Overall hydrodynamical evolution}\label{sec:overall}

\begin{figure}
	\includegraphics[width=.85\columnwidth]{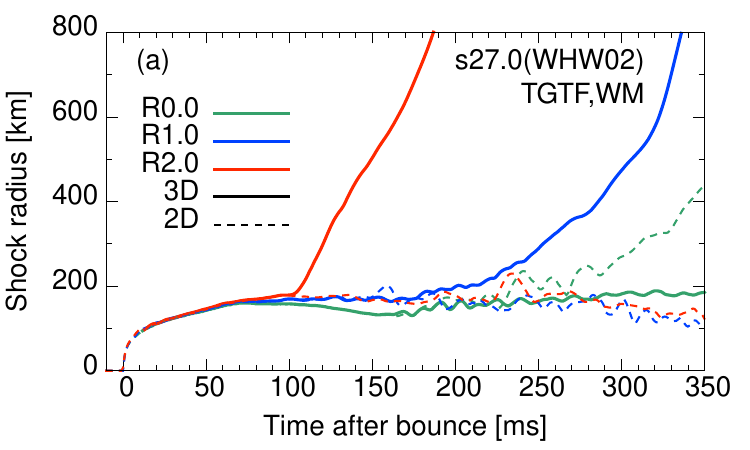}\\
	\includegraphics[width=.85\columnwidth]{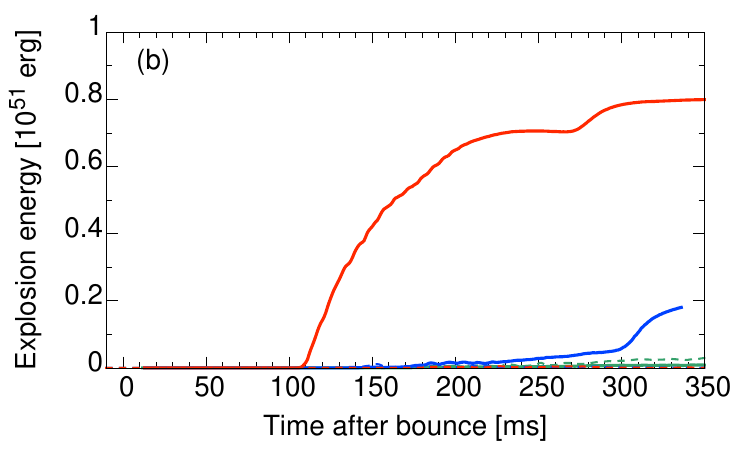}\\
	\includegraphics[width=.85\columnwidth]{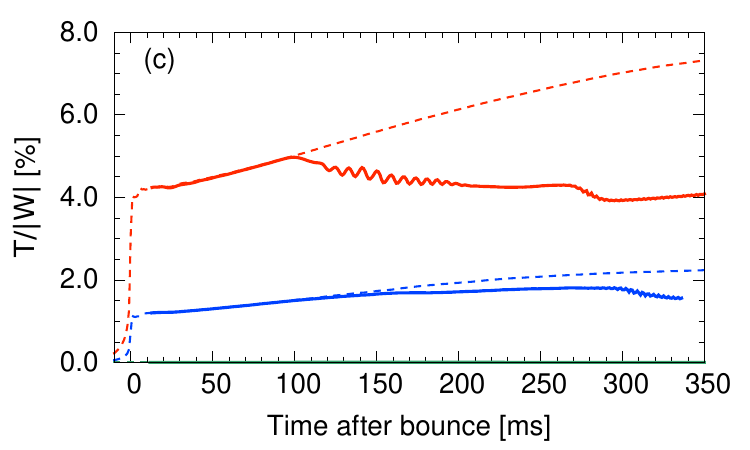}\\
	\includegraphics[width=.85\columnwidth]{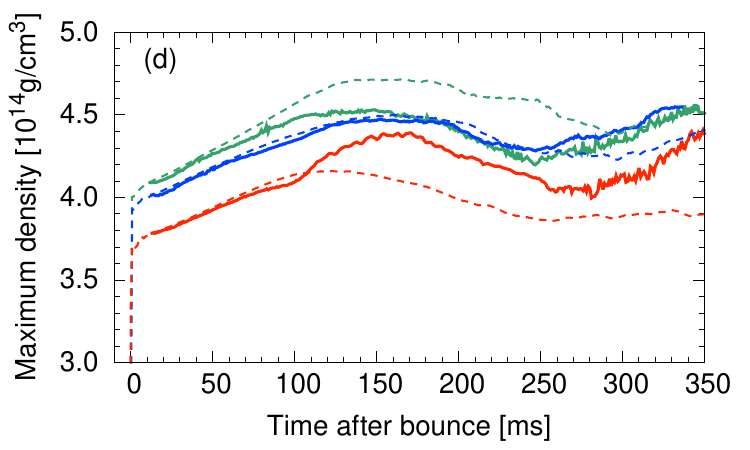}
    \caption{Key hydrodynamics quantities of our 3D models, showing the evolution of (a) the averaged shock radii, (b) the diagnostic explosion energy,  (c) $T/|W|$ (see text),  (d) the maximum density. The solid and dashed lines denote 3D and 2D models, where the color of each line represents the initial angular velocity $\Omega_0 = 0.0$ rad/s (green line), $\Omega_0 = 1.0$ rad/s (blue line), and $\Omega_0 = 2.0$ rad/s (red line).}
    \label{fig:t-Rshock;ToW;rho}
\end{figure}
We first present an overview of the hydrodynamical evolution of our 3D models. 
Figure~\ref{fig:t-Rshock;ToW;rho} shows the postbounce evolution of 
(a) the average shock radius, (b) the diagnostic explosion energy,
(c) the ratio $T/|W|$ of the rotational energy to the gravitational binding energy, (d) the maximum central density.

\begin{figure*}
\includegraphics[width=.32\linewidth]{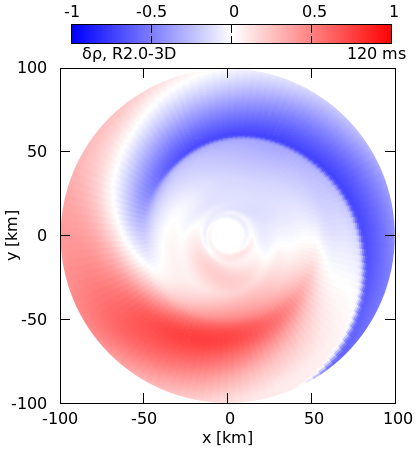}
\includegraphics[width=.32\linewidth]{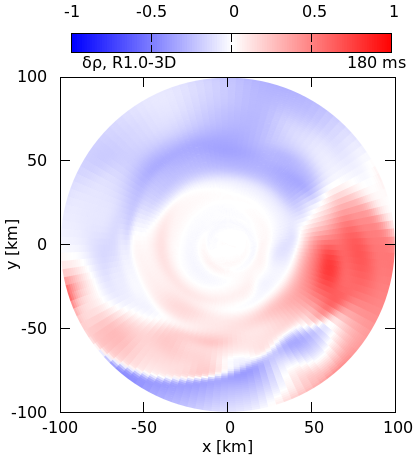}
\includegraphics[width=.32\linewidth]{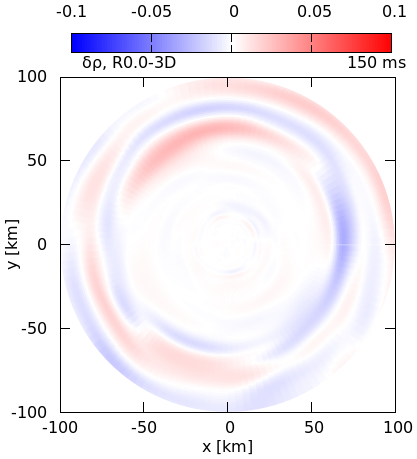}
    \caption{Density deviation in the plane $z=0$ in the models R2.0-3D (left panel), R1.0-3D (middle panel), and R0.0-3D (right panel), respectively.  The density deviation is defined by Eq.~ \eqref{eq:densitydeviation}.  The corresponding postbounce time is shown at the top right side in each panel. In the left and middle panels, the spiral structure is triggered by the \ltw instability.  In the right panel, the large scale structure is driven by the SASI.The movies of this figure is available at \url{https://doi.org/10.5281/zenodo.5463660}.}
    \label{fig:Density_z=0}
\end{figure*}

The 3D rotating models R2.0-3D, R1.0-3D explode as shown in the panel (a) of Figure~\ref{fig:t-Rshock;ToW;rho}.
The shock of R2.0-3D (red solid line) shows a clear expansion at $t_{\rm pb} \sim$ 100 ms (post bounce) and 
reaches a radius of 400 km at $\sim$ 130 ms. Following \citet{Summa16}, 
we identify the success or failure of the shock revival by considering whether the average shock radius exceeds 400 km or not.
Interestingly, the fact that the shock of R2.0-2D (red dashed line) is not revived implies that
non-axisymmetric features are crucial to revive the shock in R2.0-3D.
Similar to the rapidly rotating model R2.0-3D, the slow rotation model R1.0-3D also shows a revival of the shock (blue solid line).
The shock begins to expand at $t_{\rm pb} \sim$ 150 ms and reaches 400 km at $\sim$ 280 ms.
As discussed later, the steep revival of the shock (at $t_{\rm pb} \sim 100$ ms) in R2.0-3D and the gradual one in R1.0-3D (at $t_{\rm pb} \sim 280$ ms) are triggered by
the onset of the \ltw instability.

It is worth mentioning that the overall dynamics of the non-rotating models R0.0-3D and R0.0-2D is basically consistent with previous works with similar microphysics.
More specifically,
the 2D model (green dashed line) explodes at $t_{\rm pb} \sim$ 350ms and the 3D model (green solid line) does not explode.
This feature is consistent with \cite{Hanke13} who
employed the same progenitor \citep[s27.0 of][]{Woosley02} and a similar reaction set for neutrinos \citep[the same as in the code comparison paper of][]{OConnor18}.

The diagnostic explosion energy of the rapidly rotating model becomes $\sim 0.8\times10^{51}\,{\rm erg}$ as shown in the panel (b) of Figure~\ref{fig:t-Rshock;ToW;rho}.
Although a long-term evolution should be followed to obtain the final value of the explosion energy (e.g. \citet{Nakamura19,Bollig20}), the diagnostic energy of model R2.0-3D is close to the observed value of canonical core-collapse supernovae \citep{Tanaka09}. For our slowly rotating model R1.0-3D, the diagnostic energy ($0.2\times10^{51}\,{\rm erg}$) is shown to be smaller.
Here we use the ordinary definition for the diagnostic explosion energy:
\begin{align}
E_{\rm diag} = \int_{D}{\rm d}V \rho\left(\vec{v}^2 + \epsilon_{\rm int} +\Phi \right)   , 
\end{align}
where $V$ is the volume and $\epsilon_{\rm int}$ is the specific internal energy (see Eqs. \eqref{basic eq1} to \eqref{basic eq6} for other variables).
The integration is carried out in the region $D$ where the integrand is positive.

Since the shock revival of our R2.0-3D and R1.0-3D models is caused by the \ltw instability,
we mainly focus on these two models in what follows. The mechanism pushing the shock is basically the same as discussed in \cite{Takiwaki16}.

The starting time of the shock expansion clearly corresponds
to the onset of the non-linear evolution of the instability.
The time is $t_{\rm pb} \sim$ 100\,ms and 150\,ms for the models R2.0-3D and R1.0-3D, respectively.
After the onset, the angular momentum is extracted from the PNS and
the decrease of the centrifugal force triggers the gravitational collapse of the PNS.
As a result, the central density of the PNS becomes higher \citep{Ott05}.
The evolution of $T/|W|$ and the maximum density are shown in panels (c) and (d), respectively.
In the panel (c) at $t_{\rm pb} \sim$ 100 ms, the curve for the model R2.0 in 3D (solid red line) 
decreases and deviates from its 2D evolution (dashed red line).
In the panel (d) at $t_{\rm pb} \sim$ 100 ms, the maximum density of R2.0-3D shows a steep rise, and thereafter the evolution deviates from model R2.0-2D.
Similarly at $t_{\rm pb} \sim$ 150 ms the ratio $T/|W|$ of the slowly rotating model (R1.0-3D) decreases.
The maximum density in model R1.0-3D also increases compared to 2D at $t_{\rm pb} \sim$ 250ms although there is a significant delay with the onset of the instability.

We note from Figure~\ref{fig:t-Rshock;ToW;rho} (c) that the threshold of the \ltw instability is not merely governed by the ratio $T/|W|$: the instability begins at $T/|W|\sim 5\%$ and $2\%$ for the rapid and slow rotation models, respectively. 
 The threshold values of $T/|W|$ in our study are lower than the value measured in previous studies, such as 8\% in \cite{Ott05}. In the context of cold neutron stars,
a high degree of differential rotation is important to achieve such a low threshold value
 \citep{Shibata02,Shibata03a,Saijo03,Karino03}, but
that may not apply to hot PNS. We revisit this issue later.

\begin{figure*}
	\includegraphics[width=.32\linewidth]{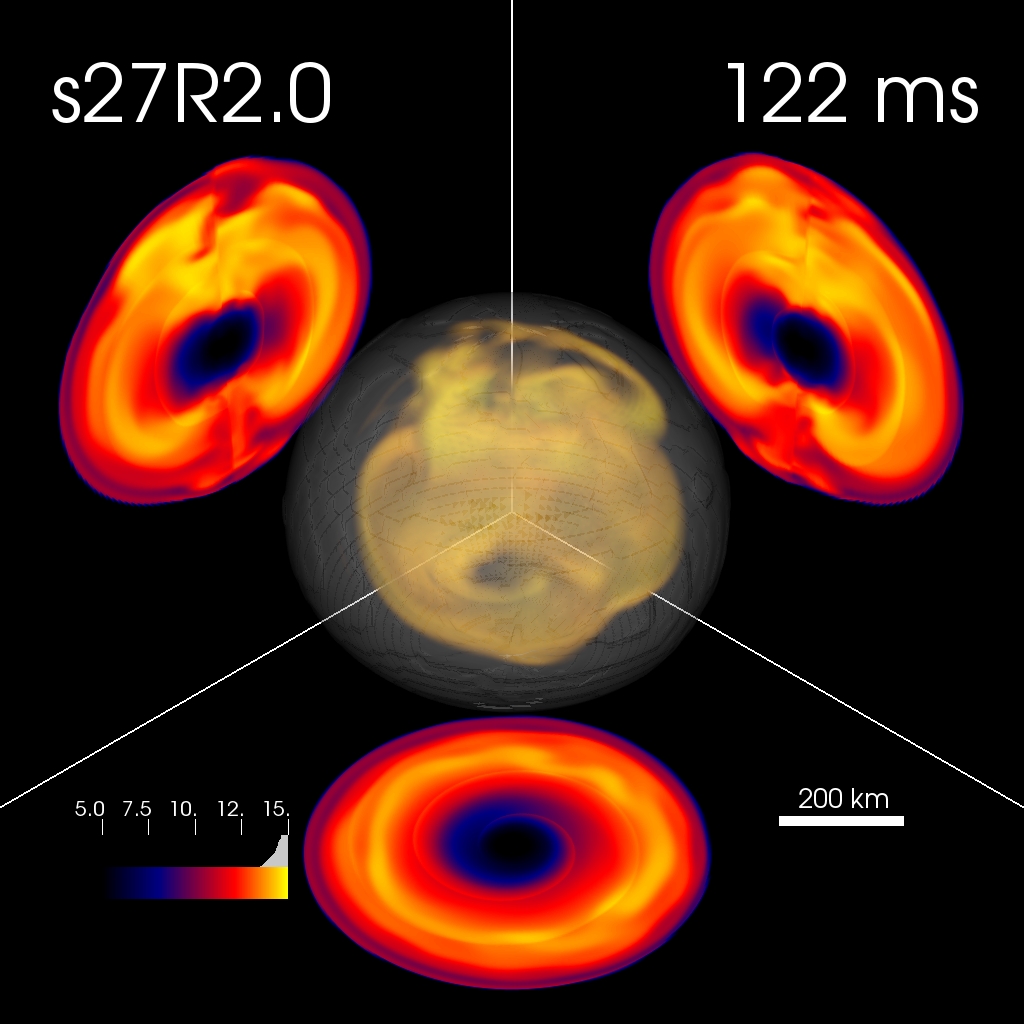}
	\includegraphics[width=.32\linewidth]{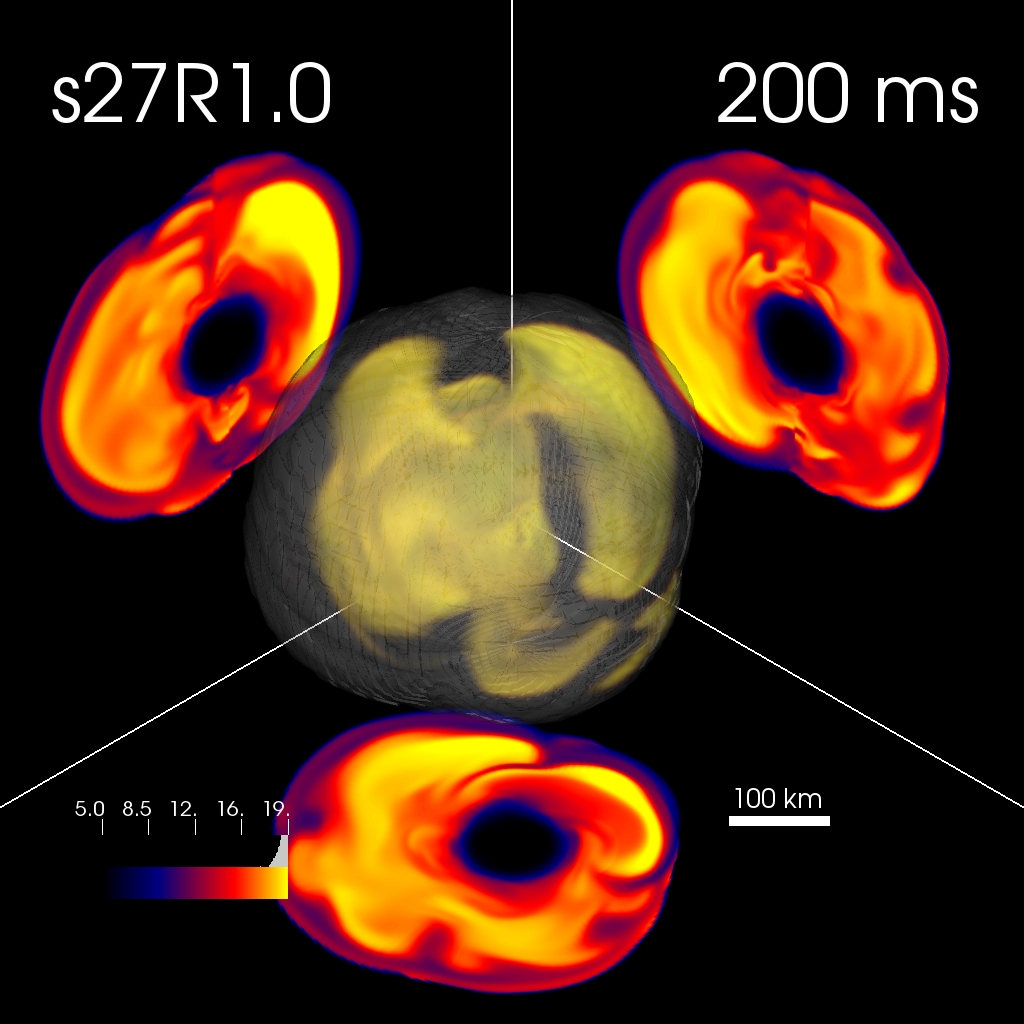}
	\includegraphics[width=.32\linewidth]{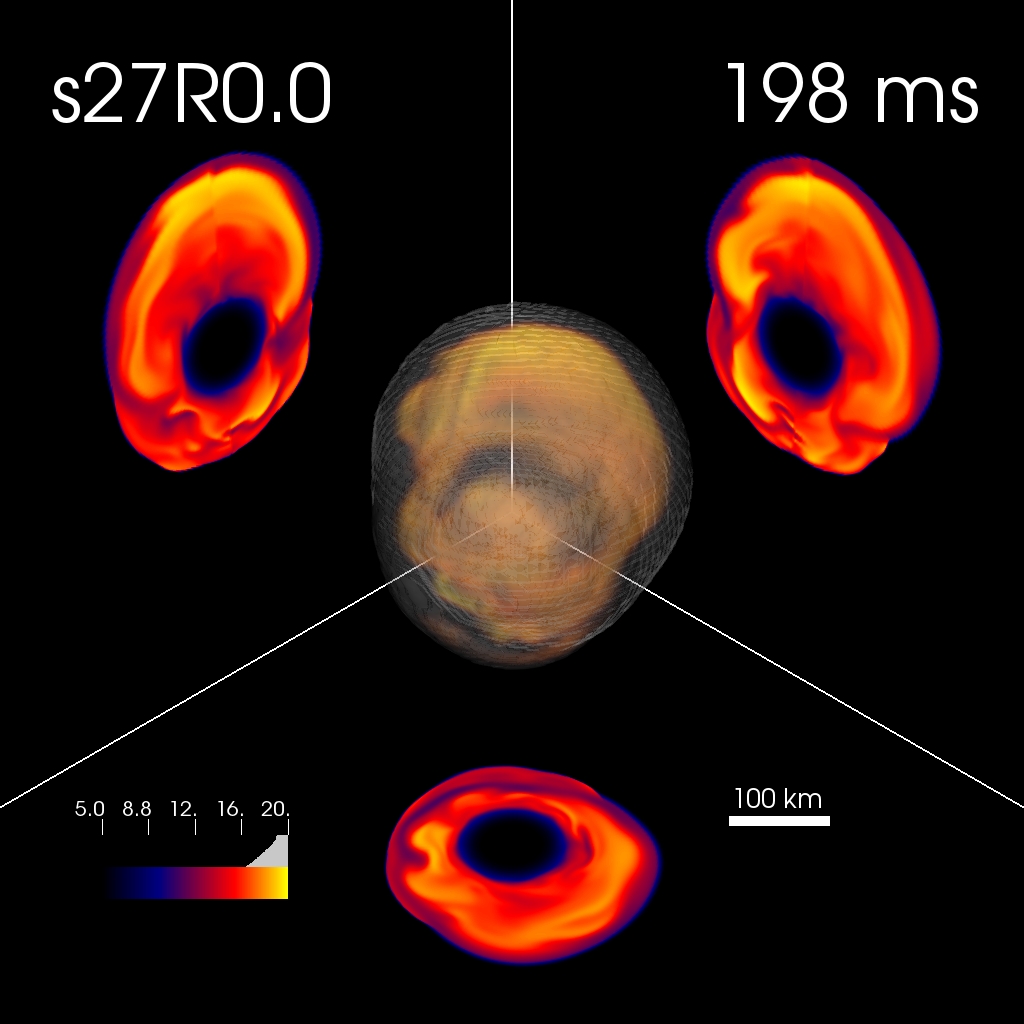}
    \caption{Structure of the shock and the entropy in our 3D models (left panel:R2.0-3D, middle panel: R1.0-3D, and right panel: R0.0-3D) at representative postbounce times (as given in the upper right corner of each panel). The central object in each panel shows a volume-rendered image of the entropy. The outermost whitish sphere depicts the position of the shock. The images on the cube walls show entropy distributions (the color coding as given by the color bars in the lower left corners with black, blue, and green signaling low values) in the $x$-$y$, $x$-$z$, and $y$-$z$ planes. A scale stick in each panel gives a measure of the size. The movies of this figure is available at \url{https://doi.org/10.5281/zenodo.5463660}.}
    \label{fig:3dcolormap}
\end{figure*} 

The features of the instability clearly appear in the density profile. Figure~\ref{fig:Density_z=0} shows a representative time snapshot of the density deviation at $z=0\ (\theta=\pi/2)$ plane, which is defined by
\begin{equation}
    \delta \rho \equiv
     \left. \frac{\rho -\langle \rho \rangle}{\langle \rho \rangle} \right|_{\theta=\frac{\pi}{2}}, \label{eq:densitydeviation}
 \end{equation}
 where 
 \begin{equation}
 \langle \rho \rangle \equiv
    \left. \frac{1}{2\pi}\int{\rm d}\phi\,\,\rho(r,\theta,\phi) 
    \right|_{\theta=\frac{\pi}{2}}.
\end{equation}

The spiral structure in density is a typical feature of the \ltw instability \citep{Ott05}. 
The left panel of Figure~\ref{fig:Density_z=0} shows the density deviation for model R2.0-3D.
The spiral arm of $m=1$ is seen very clearly.
The color bar spans from $-1.0$ to $1.0$. The amplitude of the deviation in this mode is large and comparable to the average .
This profile is similar to that shown in the top panel of Figure 3 in \cite{Takiwaki16}.
The middle panel of Figure~\ref{fig:Density_z=0} is for model R1.0-3D. The profile is clearly non-axisymmetric but the flow patterns look more disturbed compared to the left panel that exhibits a clear spiral deformation. 
As will be shown later, both $m=1$ and $m=2$ modes develop in this model, which results in making the spiral feature less pronounced. The right panel of Figure~\ref{fig:Density_z=0} is for model R0.0-3D, which shows a large-scale, rather concentric structure that is produced by the SASI as discussed later. Note in this panel that the amplitude of the deviation is much smaller than that of the \ltw instability (see the color bar, spanning from $-0.1$ to $0.1$ in the right panel).

Intriguing features of the instability
(the \ltw instability or the SASI) also appear in the shape of the shock.
Figure~\ref{fig:3dcolormap} shows the 3D structure of the shock and the entropy distribution.
In each panel, the volume-rendered entropy is surrounded by a whitish shell that represents the position of the shock.
The images on the cube walls show the entropy distribution in the $x$-$y$, $x$-$z$, and $y$-$z$ planes.

The entropy structures reflect the non-linear dynamics after the onset of the instability.
The spiral pattern similar to the previous figure is also visible in the inner region of the entropy contour in the bottom cube wall of the left panel.
The spiral arm colored in red expands from the PNS (low-entropy, colored by blue) to the outer region.
In the outer region, the high entropy region appears as a yellow ring in the bottom cube wall. It is produced by the propagation of the spiral perturbation. The right panel of Figure~\ref{fig:3dcolormap} illustrates the strong (sloshing) SASI activity for the non-rotating model R0.0-3D.

The bottom cube wall in the middle panel of Figure~\ref{fig:3dcolormap} illustrates the highly non-axisymmetric feature in model R1.0-3D. However the shape of the shock is not simply ascribed to the $m=1$ or $m=2$ deformation, in contrast to model R2.0-3D (the left panel). Neutrino-driven convection and the SASI are most likely to explain the vigor of the shock deformation for this model.
This model might be viewed as an intermediate model between the \ltw instability-dominated model (left panel) and the SASI-dominated model (right panel).
A parametric study like \cite{Kazeroni17} should be done also in the context of self-consistent simulations, in order to understand the impact of the initial distribution of angular momentum in the core on the dynamical evolution in the postbounce phase, dominated by the SASI or the \ltw instability. 

\subsection{The Low-T/|W| instability in the accreting PNS}\label{sec:detailofTW}

Contrary to its name,
the \ltw instability is {\it not} triggered merely 
by the value of $T/|W|$ in our simulations (see panel (c) of Figure~\ref{fig:t-Rshock;ToW;rho}).
Though the nature of the instability has not been completely understood,
several key ideas have been proposed to give insights 
into its mechanism.
In Section~\ref{sec:cororegion}
we follow the same type of analysis as performed in cold neutron
stars using the concepts of pattern frequency and corotation radius.
In Section~\ref{sec:rossby}, we propose a new scenario for the \ltw instability based on the destabilization of Rossby waves near the convective boundary, where stratification does not prevent the destabilizing effect of differential rotation. 
In Section~\ref{sec:sasispiral}, we briefly comment on the relation between the spiral SASI and the \ltw instability.

\subsubsection{The corotation instability}\label{sec:cororegion}
We start with a similar analysis as in cold neutron stars.
According to \citet{Watts05} and the linear analysis of \cite{Passamonti15}, the \ltw instability is triggered if the pattern frequency of the $f$-mode, $\sigma_f$, 
matches the angular frequency $\Omega$ of the fluid at some radius $r_{\rm co}$ called the corotation radius:
\begin{align}
\sigma_f &= \Omega(r_{\rm co}).\label{eq:sigmaf-corot}
\end{align}

In a differentially rotating star, the fluid angular frequency reaches a maximum $\Omega_{\rm max}$ at the center and a minimum $\Omega_{\rm min}$ at the edge. The necessary condition of the \ltw instability is expected if the $f-$mode pattern frequency
is intermediate between the maximum and minimum angular frequencies:
\begin{align}
    \Omega_{\rm min} < \sigma_f < \Omega_{\rm max},\label{eq:Pcondition}
\end{align}
which we refer to as the Watts condition.

\begin{figure}
	\includegraphics[width=.9\columnwidth]{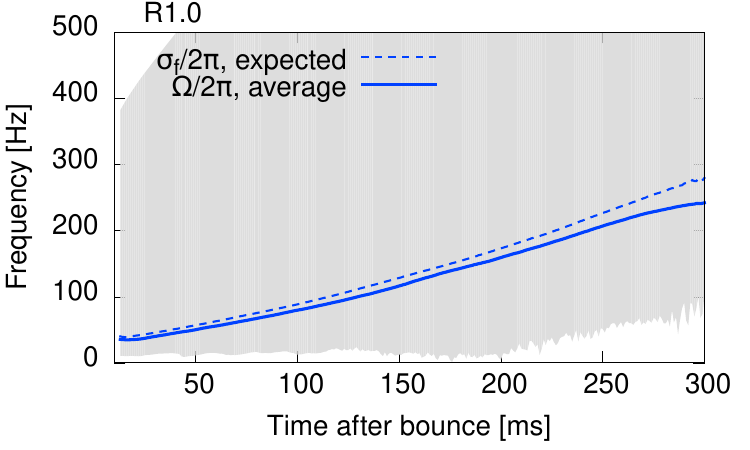}\\
	\includegraphics[width=.9\columnwidth]{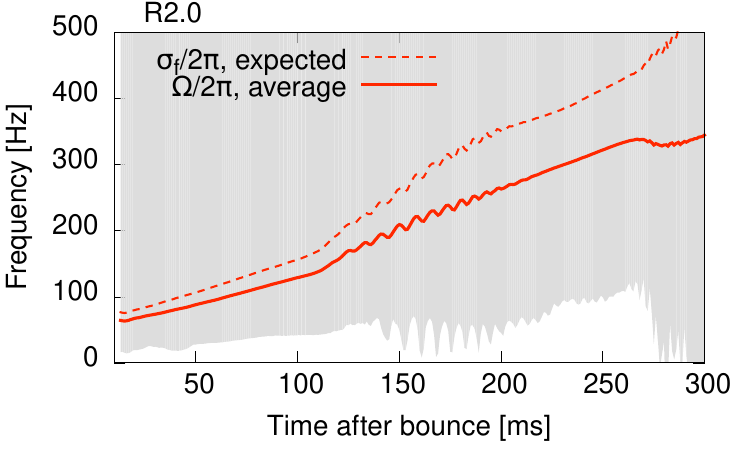}
    \caption{Time evolution of the corotation region (shaded region), the average angular frequency ($\Omega_{\rm ave}$) in the PNS,  and the expected $f$-mode pattern frequency. See Eqs. \eqref{eq:omgmax} to \eqref{eq:omgave} and the following text for the definition. Note that the maximum angular frequency is much higher that 500\,Hz.}
    \label{fig:t-Omega}
\end{figure}

We first discuss whether the Watts condition is satisfied or not in our models using Figure~\ref{fig:t-Omega}.
For this, we define the following variables:
\begin{align}
    \Omega  &= v_\phi/r_\perp,\\
    \Omega_{\rm max} &= \max \left.[\Omega]\right._{\rm ePNS},\label{eq:omgmax}\\
    \Omega_{\rm min} &= \min \left.[\Omega]\right._{\rm ePNS},\label{eq:omgmin}\\
    \Omega_{\rm ave} &= 
    \int_{\rm ePNS}\rho v_{\phi} r_{\perp}{\rm d}V 
    \bigg/
    \int_{\rm ePNS}\rho r_{\perp}^2{\rm d}V, \label{eq:omgave}
\end{align}
where $v_\phi$ is the velocity in the $\phi$ direction and $r_\perp$ is the distance from the rotation axis.
The integral in Eq.~\eqref{eq:omgave}, and the maximum and minimum operators are restricted to the region inside the PNS, defined by the fiducial density of $\rho > 10^{11}\,{\rm g/cm^3}$.
To focus on the properties in the equatorial plane, we additionally limit the region as $60^{\circ} < \theta <120^{\circ}$ in the analysis.
The subscript "ePNS" in the equations refers to this restricted region.
In Eq.~\eqref{eq:omgave}, we follow the definition of the averaged angular velocity in \cite{Ott06}.
The corotation region (gray region) in the Figure~\ref{fig:t-Omega} is defined as $\Omega_{\rm min}/2\pi < f <\Omega_{\rm max}/2\pi$.
To obtain a rough estimate of the $f$-mode frequency of the PNS, we employ the empirical formula of \cite{Sotani19}:
$f_f = -87.34 + 4080.78\left(\frac{M_{\rm PNS}}{1.4M_\odot}\right)^{1/2}
    \left(\frac{R_{\rm PNS}}{10\,{\rm km}}\right)^{1/2}[{\rm Hz}]$, where $M_{\rm PNS}$ and $R_{\rm PNS}$ denote the mass and radius of the PNS.
    Since there is no analytical formulae to estimate the $f$-mode frequency, we use this formula only as an indicative reference remembering that it was constructed to fit the $l=2$, $m=0$ perturbations of a non-rotating PNS models \citep{Sotani19}. Keeping these caveats in mind, we proceed to estimate the pattern frequency as $\sigma_f/2\pi = f_f/m$ for $m\neq 0$. 
     In Figure~\ref{fig:t-Omega}, 
      the pattern frequency is estimated as $\sigma_f = f_f/2$ for model R1.0-3D (top panel) and 
$\sigma_f = f_f$ for model R2.0-3D  (bottom panel).
Here we choose $m$ to adjust the real pattern frequency, e.g. $m=1$ for model R2.0-3D (e.g. the left panel of Figure~\ref{fig:Density_z=0}) and $m=2$ for model R1.0-3D (as discussed later). 

The Watts condition defined by Eq.~\eqref{eq:Pcondition} is always satisfied in our rotating models.
As shown in Figure~\ref{fig:t-Omega},
the gray region in the figure is the range of rotation frequencies and 
the dashed line is the expected pattern frequency.
The pattern frequencies are always in the range of the rotation frequencies.
Note that $\Omega_{\rm max}$ is much higher than 500\,Hz and $\Omega_{\rm min}$ is smaller than 100\,Hz 
even if the averaged $\Omega$ spans 100--300 Hz from 100\,ms to 300ms postbounce.
The broad range of the rotation frequencies indicates a strong differential rotation.
Our Figure~\ref{fig:t-Omega} is similar to the right panel of Figure 3 in \cite{Passamonti15}, which corresponds to the model with the highest differential rotation in their study. 
Even though our estimation of the $f$-mode is not accurate as mentioned above, we expect the true pattern frequency to lie within the wide range of rotation frequencies.

\paragraph{Slowly rotating model}\label{sec:slowlyrotatingmodel}
The evaluation of the pattern frequency and the corotation radius is necessary to reach a deeper understanding of the instability. We analyze the R1.0-3D model before the R2.0-3D model.

\begin{figure}
	\includegraphics[width=\columnwidth]{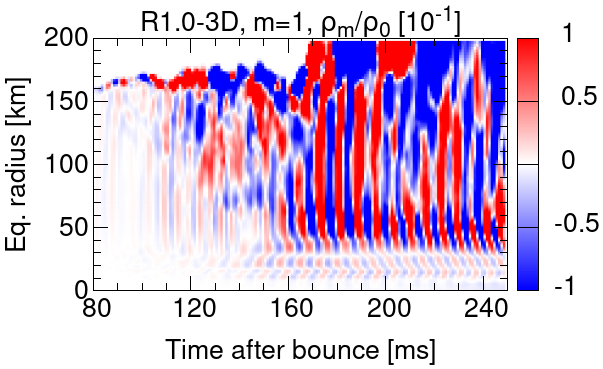}\\
	\includegraphics[width=\columnwidth]{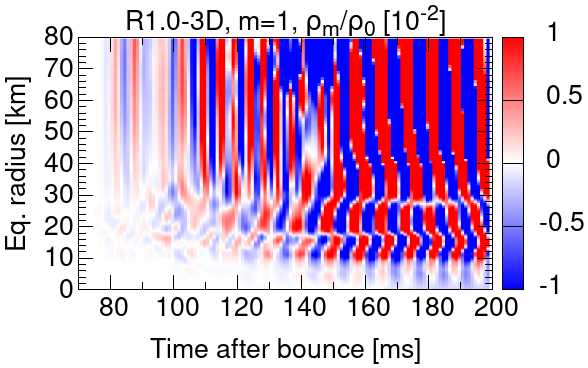}\\
	 \caption{The top panel is the space-time evolution of the $m$=1 density deviation $\rho_1/\rho_0$ (see text for the definition) of model R1.0-3D. The $y$ axis represents the equatorial radius. The bottom panel is similar to the top panel, but showing a zoom up in the central region (note the difference of the $y$ scale).}
    \label{fig:R103Dt-r-d}
\end{figure}

To extract the spiral motion of the instability, we define a harmonic decomposition of density as follows:
\begin{align}
\rho_{m}(t,r) =& \int{\rm d}\phi\,\rho(t,r,\theta=\pi/2,\phi) \cos(m\phi).\label{eq:rho-hamonics}
\end{align}
The relative component, $\rho_m/\rho_0$, for $m=1$ is shown as a function of time and equatorial radius in Figure~\ref{fig:R103Dt-r-d}.
A spiral component appears as an oscillation of $\rho_{m}(t,r)$ propagating to the outer radius.

In Figure~\ref{fig:R103Dt-r-d}, the stripe pattern corresponds to the strong $m=1$ oscillation.
This pattern is produced at the bottom of the stripe (at a radius above $\sim$ 20--30\,km, bottom panel) and propagates to the outer region \citep[see also the right panel of (a) in Figure 3 of][]{Takiwaki16}.
The clear stripe pattern appears after $t_{\rm pb} \sim 150\,{\rm ms}$, which approximately corresponds to the transiton from the linear to non-linear phase of the instability as discussed later.
The pattern propagates to $\sim$ 150--200 km
(the top panel).
The density deviation for the mode $m=2$ has a similar stripe pattern.

\begin{figure}
	\includegraphics[width=\columnwidth]{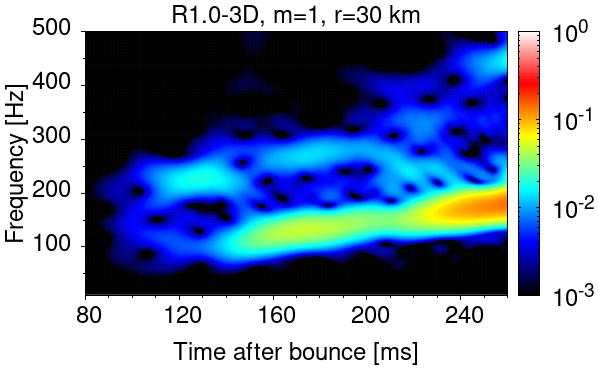}\\
	\includegraphics[width=\columnwidth]{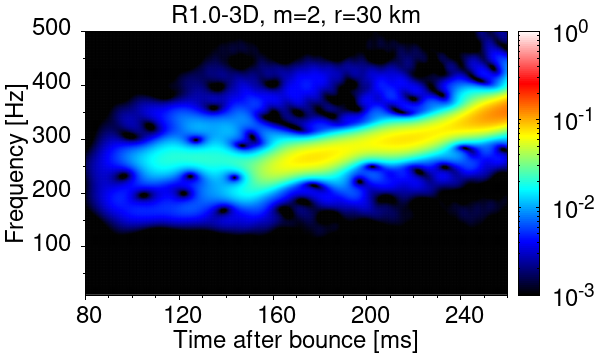}
	    \caption{ Top: Spectrogram of the $\rho_1/\rho_0$ at 30km.
             Bottom: Spectrogram of the $\rho_2/\rho_0$  at 30km. }
    \label{fig:R103Dt-f-d}
\end{figure}

In order to evaluate the frequency of the pattern,
we compute spectrograms of $\rho_m/\rho_0$ fixing $m$ and $r$.
The spectrograms for $m=1$ and $m=2$ for $r=30\,{\rm km}$ are shown in the top and bottom panels of Figure~\ref{fig:R103Dt-f-d}, respectively. 

A strong oscillation of the mode $m=1$ appears at 150\,ms (Figure~\ref{fig:R103Dt-f-d}, top panel).
We refer to this as the onset of the non-linear phase.
Later on, the frequency of the mode increases as a function of time.

In the spectrogram for $m=2$ (Figure~\ref{fig:R103Dt-f-d}, bottom panel), one can see an oscillation at 250\,Hz at 90--150\,ms, which is before the onset of the strong oscillation. We refer to this phase as the linear phase, where the pattern is visible but does not strongly affect the overall dynamics. 

It is important to point out that 
 the frequency of the mode $m=2$ (bottom panel
  of Figure~\ref{fig:R103Dt-f-d}) is twice as large as the mode $m=1$ (top panel): $f_{{\rm mode},2} \sim 2 f_{{\rm mode},1}$. Since the power excess in the spectrogram (Figure~\ref{fig:R103Dt-f-d}) increases approximately linearly with time for both the $m = 1$  and $m = 2$ modes, we fit the mode frequencies as a function of the postbounce time ($t_{\rm pb}$ [ms]) as follows:
\begin{align}
f_{{\rm mode},1} &\sim
\begin{array}{cc}
 0.6 ( t_{\rm pb} -150)+110\,{\rm Hz}&  (150\,{\rm ms}< t_{\rm pb})   \label{eq:f1inR1.0}\\
\end{array},\\
f_{{\rm mode},2} &\sim
\begin{cases}
250 \,{\rm Hz}& (90\,{\rm ms} < t_{\rm pb} < 150\,{\rm ms}) \\
1.2 (t_{\rm pb}-150)+220 \,{\rm Hz}& (150\,{\rm ms} < t_{\rm pb})  \label{eq:18}
\end{cases}.
\end{align}
With these two fitting formulae we can evaluate the pattern frequency $\sigma_{\rm pat}$, and finally find the co-rotation radius $r_{\rm co}$ via Eq.~(\ref{eq:sigmaf-corot}):
\begin{align}
 2\pi f_{{\rm mode},m}/m &= \sigma_{\rm pat} = \Omega(r_{\rm co}).\label{eq:sigma-corot}
\end{align}
Here $\sigma_{\rm pat}$ is determined by the Fourier analysis of the harmonic decomposition of the density.
The $f-$mode was considered in Eq.~\eqref{eq:sigmaf-corot} but
the mode in Eq.~\eqref{eq:sigma-corot} is not necessarily the $f-$mode.
Note that the pattern frequency does not depend on $m$ since $\sigma_{\rm pat}/2\pi  = f_{{\rm mode},2} /2 = f_{{\rm mode},1}$ in the non-linear phase (i.e., at $t_{\rm pb} > 150\,{\rm ms}$).

\begin{figure}
	\includegraphics[width=0.99\columnwidth]{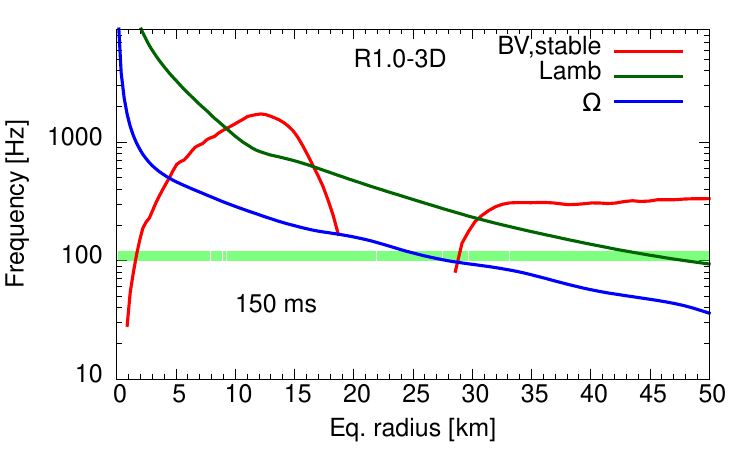}\\	
	\includegraphics[width=0.99\columnwidth]{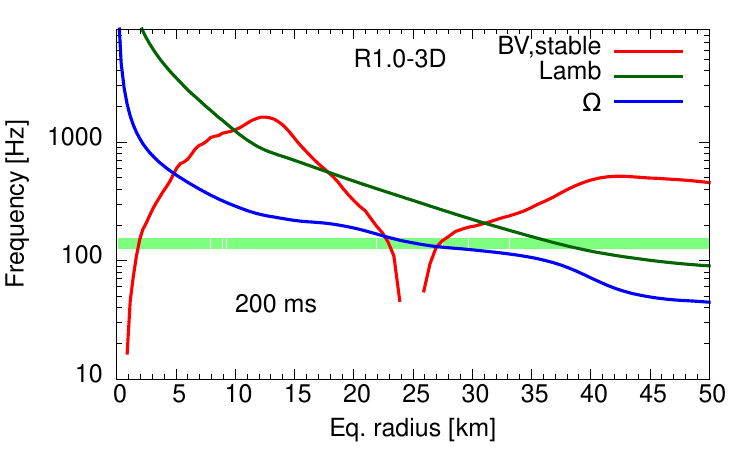}
    \caption{The propagation diagram for our R1.0-3D model. Top: At 150ms, the Green band is 110\,Hz. Bottom: At 200ms, the Green band is 135\,Hz.
    }
    \label{fig:R1.0-3Dpropagation}
\end{figure}

The time evolution of the corotation radius is depicted in Figure~\ref{fig:R1.0-3Dpropagation}.
The top panel shows the radial dependence of reference frequencies at $t_{\rm pb} = $ 150\,ms.
The pattern frequency is displayed as a green band, and the averaged fluid angular frequency $\Omega$ is shown as the blue line as a function of the equatorial radius.
The two lines cross at $\sim 26$km, which defines the corotation radius.
A different snapshot is shown at 200ms in the middle panel, where the corotation radius is moved to $\sim 25$km.
The time evolution of the corotation radius is shown as the blue line in the bottom panel.
The corotation radius decreases after 150 ms.
This shrinking motion suggests an increase of the mode frequency.
Those features are similar to those in \cite{Shibagaki20}.

According to the analysis of \cite{Yoshida17} in the context of cold neutron stars (NSs),
the \ltw instability is triggered by a stellar oscillation trapped between the corotation radius and the NS surface. The frequency of the stellar oscillation should match the pattern frequency.
The three oscillation modes existing in a star are called $p-$, $g-$, and $f-$modes.
The propagation diagram is useful  to identify the class of the mode.
In the diagram, the radial profiles of the \BV frequency and the Lamb frequency are defined as follows:
\begin{align}
    \Omega_{\rm BV}^2  =& -\frac{1}{\rho}\frac{\partial \Phi}{\partial r}
    \left(\frac{1}{c_s^2}\frac{\partial p}{\partial r} - \frac{\partial \rho}{\partial r}\right),\\
    \Omega_{\rm Lamb} =& \frac{\sqrt{\ell(\ell+1)}c_s}{r},
\end{align}
where $p$ is the pressure. 
The allowed region for wave propagation is visible on the diagram.
$p-$modes can propagate with frequencies higher than $\Omega_{\rm Lamb}$.
$g-$modes can propagate with frequencies lower than $\Omega_{\rm BV}$ if $\Omega_{\rm BV}^2$ is positive.
In the top and bottom panels of Figure~\ref{fig:R1.0-3Dpropagation},
$\Omega_{\rm BV}$ and $\Omega_{\rm Lamb}$ are over-plotted on the angular frequency and the pattern frequency
at 150\,ms and 200\,ms, respectively.
The Lamb frequency displayed corresponds to the lowest $\ell=1$ mode.

One remarkable property can be pointed out from the panels:
the position of the convection layer in the PNS coincides with the corotation radius.
The negative lepton gradient is responsible for the convection inside the PNS  \citep[e.g.][]{Buras06b}.
In the top (bottom) panel, $\Omega_{\rm BV}$ is disconnected at 18-28\,km (24-26\,km)
and the region corresponds to the convective layer since the
\BV frequency becomes an imaginary number there.
Interestingly, the position of the convective layer matches the corotation point (the crossing point of the blue line and the green band). 
From $\sim 100$\,ms to $\sim 300$\,ms, the position of the convective layer is almost the same as the corotation radius.
The role of the convection on the \ltw instability is discussed in Section \ref{sec:rossby}.

\begin{figure}
	\includegraphics[width=\columnwidth]{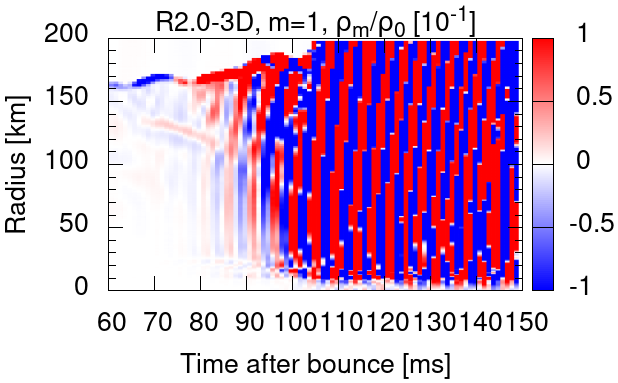}\\
	\includegraphics[width=\columnwidth]{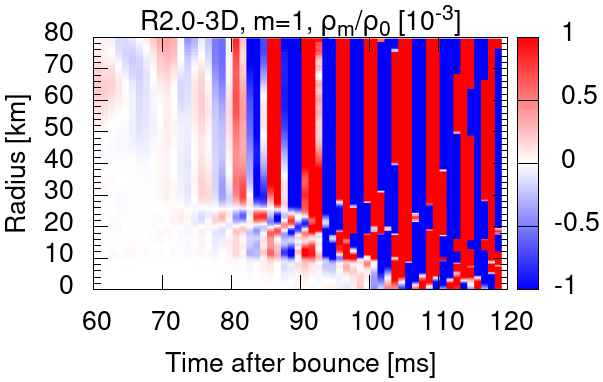}
	 \caption{Top: Time evolution of $m$=1 density deviation, $\rho_1/\rho_0$ for model R2.0-3D.
             Bottom: Zoom up of the top panel as similar to Figure~\ref{fig:R103Dt-r-d}.}
    \label{fig:R203Dt-r-d}
\end{figure}

\begin{figure}
	\includegraphics[width=\columnwidth]{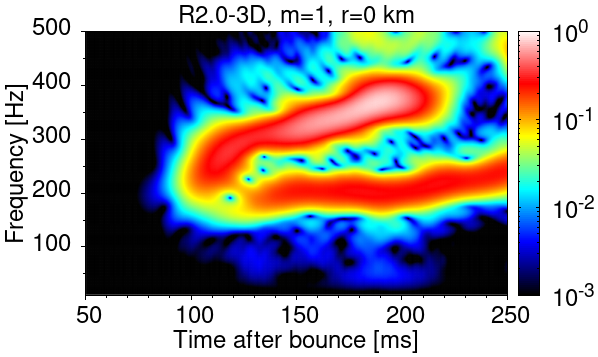}
    \caption{Spectrogram of the $\rho_1/\rho_0$ at 30km for model R2.0-3D.}
    \label{fig:R20t-f-pat}
\end{figure}

\begin{figure}
	\includegraphics[width=0.99\columnwidth]{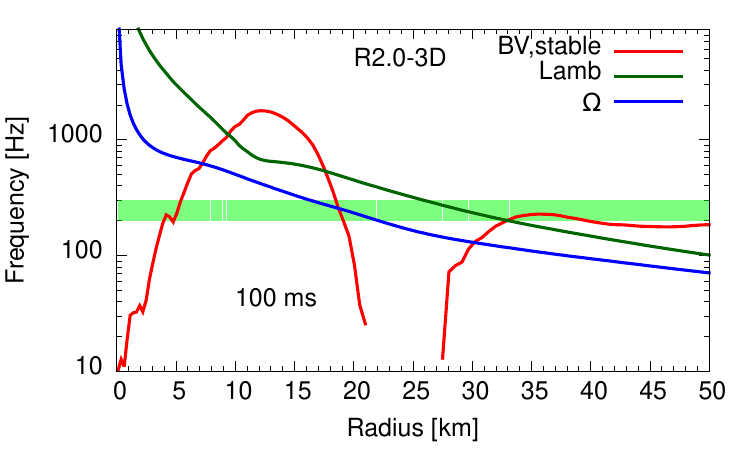}\\
	\includegraphics[width=0.99\columnwidth]{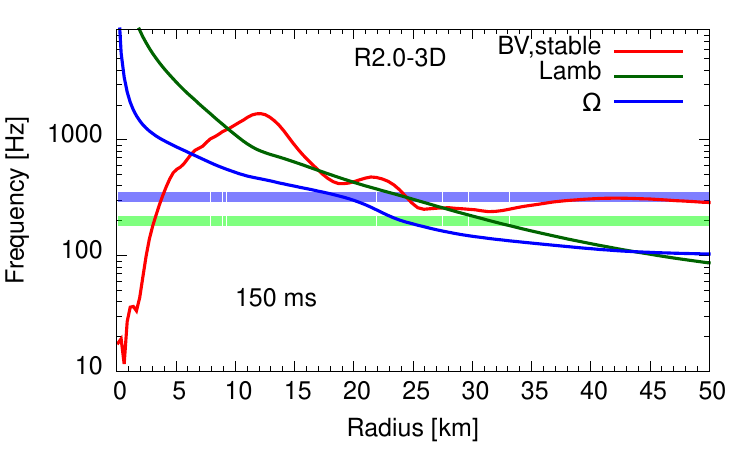}
    \caption{ Top: Propagation diagram at $t_{\rm pb} =$ 100\,ms. Green band is 200-300\,Hz.
             Middle: Propagation diagram at 150ms. Green band: 200\,Hz. Purple band: 320\,Hz.
             }
    \label{fig:R2.0-3Dpropagation}
\end{figure}

\paragraph{Rapidly rotating model}\label{sec:rapidlyrotatingmodel}

We apply the same analysis used in the previous section to the strong rotation model R2.0-3D although the hydrodynamic behavior is more complicated.
Again the mode $m=1$ is extracted using Eq.~\eqref{eq:rho-hamonics}.
In this model, the amplitude of the mode $m=2$ is sufficiently smaller than $m=1$ to be ignored.
The space-time diagram in Figure~\ref{fig:R203Dt-r-d} shows 
the propagation of the mode $m=1$.
The mode originates from the deformation of the PNS.
The overall feature is shown in the top panel and the detailed feature near the PNS is shown in the bottom panel.
As one can see in the bottom panel, the \ltw begins at 80 ms at 20 km.
The amplitude of the mode immediately becomes large and propagates to the outer region.
At 90 ms, it reaches the shock and pushes it (see the top panel).

The mode frequency of Eq.~\eqref{eq:sigma-corot} in the rapid rotation model is shown in Figure~\ref{fig:R20t-f-pat}.
In this model, we found two modes: one has a frequency that ramps up from 200\,Hz to 400\,Hz during $t_{\rm pb} \sim$ 100 - 200 ms
and the other has a relatively stationary frequency of 200\,Hz.
Note that both are $m=1$ modes.
The higher frequency mode is not the overtone of the lower frequency mode since their time evolutions are significantly different. Again, the evolution of the mode frequency is fitted (partly) by using the linear function of $t_{\rm pb}$ as,
\begin{align}
f_{{\rm mode},1,{\rm low}} &=
\begin{array}{cc}
 200\,{\rm Hz}&  (80\,{\rm ms}< t_{\rm pb})   \label{eq:f_lowinR2.0}\\
\end{array},\\
f_{{\rm mode},1,{\rm high}} &= 
\begin{array}{cc}
 t_{\rm pb}+170\,{\rm Hz}&  (100\,{\rm ms}< t_{\rm pb}).   \label{eq:f_hghinR2.0}\\
\end{array}
\end{align}
We refer to $t_{\rm pb} \sim$ 80--100 ms as the linear phase and the evolution after 100 ms as the non-linear phase.

From the two pattern frequencies, we can estimate two corotation radii.
At $t_{\rm pb} \sim$ 100\,ms, the frequencies of the two modes are almost the same and the corotation radius is located at a radius of $\sim$ 20 km.
In the top panel of Figure~\ref{fig:R2.0-3Dpropagation}, the angular frequency (blue line)
crosses the pattern frequency (green band) at 20\,km.
At 150\,ms, on the other hand, there are two pattern frequencies and two corotation points.
In the bottom panel of Figure~\ref{fig:R2.0-3Dpropagation},
the two pattern frequencies (green and purple band)  
cross the angular frequency at 23\,km and 18\,km, respectively.

The propagation diagram is also over-plotted in the top and bottom panels of Figure~\ref{fig:R2.0-3Dpropagation}.
The position of the convective layer also roughly matches the two corotation radii.
In the top panel, the \BV frequency becomes imaginary from 21\,km to 27\,km.
In the bottom panel, it is hard to identify the convective layer since the negative $Y_e$ gradient is flattened by the non-linear effect. Though the \BV frequency is not imaginary, two dips are found at 18\,km and 25\,km.
The convective layer is in this vicinity.

\begin{figure*}
	\includegraphics[width=0.3\linewidth]{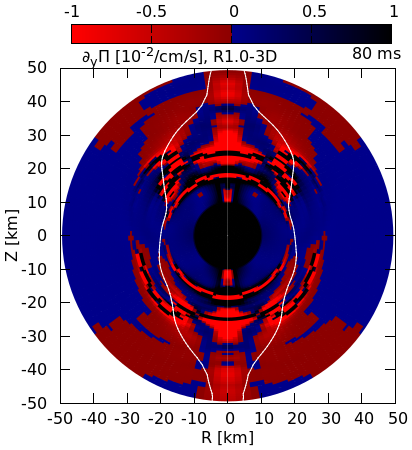}
	\includegraphics[width=0.3\linewidth]{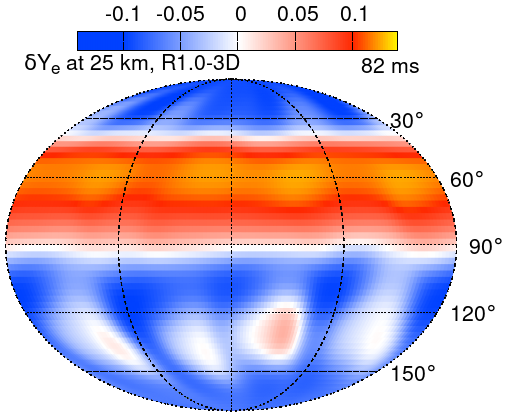}
	\includegraphics[width=0.3\linewidth]{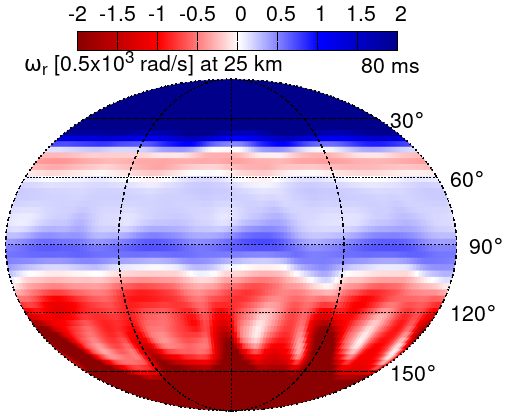}\\
	\includegraphics[width=0.3\linewidth]{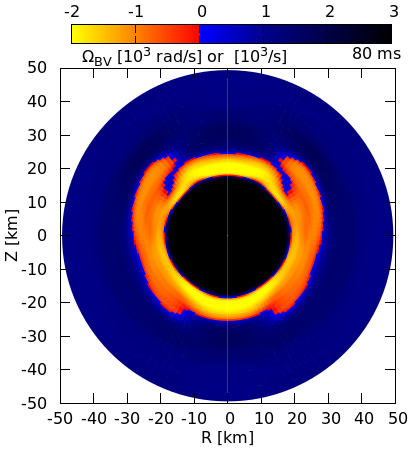}
	\includegraphics[width=0.3\linewidth]{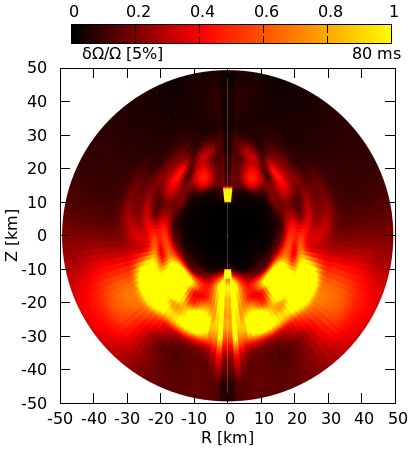}
	\includegraphics[width=0.3\linewidth]{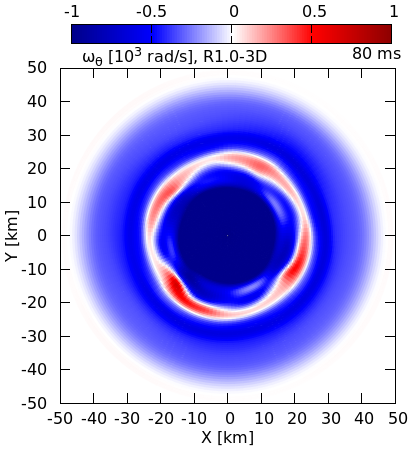}
    \caption{Snapshots of variables at $\sim 80\,{\rm ms}$ in R1.0-3D model.
    {\bf Top left}: y-derivative of the potential vorticity, $\partial_y\Pi$ in Eq.~\eqref{eq:instability-criterion}.To calculate it, we use azimuthal averaged quantities. $R,Z$ are the radius and height of cylindrical coordinates, respectively. The white line corresponds to the corotation radius of $\Omega/2\pi$ = 125Hz.
    {\bf Bottom left}: \BV frequency. As the top left panel, the azimuthal averaged quantities are used to calculate the frequency.
    {\bf Middle top}: $\delta Y_e/\langle Y_e\rangle_{\theta\phi}$ at $r=25\,{\rm km}$ in Mollweide projection, where $\delta Y_e=Y_e - \langle Y_e\rangle_{\theta\phi}$. Here the bracket means the angle and azimuthal average.
     {\bf Right top}: r-component of vorticity, $\omega_r$ at $r=25\,{\rm km}$ in Mollweide projection.
     {\bf Right bottom}: $\theta$-component of vorticity, $\omega_\theta$ plotted in $z=0$ plane.
     {\bf Middle bottom}: the dispersion of angular frequency, $\delta \Omega/\langle \Omega \rangle_{\phi}$, is plotted in $R$-$Z$ plane. Here $\delta \Omega= \sqrt{\langle ( \Omega - \langle \Omega \rangle_\phi)^2 \rangle_{\phi}}$, where the bracket means azimuthal average.
    In all plots, we use some appropriate color for $r<10\,{\rm km}$ where we assume radial advection.}
    \label{fig:R1.0-3D-ss-80ms}
\end{figure*}

\subsubsection{Rossby wave instability}\label{sec:rossby}
The previous section has revealed the complex nature of the \ltw instability in the PNS and its relation to differential rotation and the convective layer.
In this section, we take a different approach and evaluate the criterion for shear instabilities in stratified and unstratified gases.
Since this idea is significantly different from the ideas in the literature of the \ltw instability in isolated neutron stars,
we explain the basic concept following \cite{Spruit84} and \cite{Lovelace14}.

In the context of planet formation, \citet{Lovelace99,Li00} have established
that the non-axisymmetric instability of a protoplanetary disk is associated to 
a trapped Rossby wave \citep[see][for a quick review]{Lovelace14}. Their criterion of the instability
has been also tested in stellar interiors by \cite{Ou06}.

\cite{Spruit84} considers a similar instability in the geometry of stars  under the quasi-geostrophic approximation though they did not call it Rossby wave instability. They start from the propagation equation of the Rossby wave in a star:
\begin{equation}
    \left(\partial_t +\Omega\partial _\phi\right)\left[
    \Delta_{\theta\theta} \psi
    +\Delta_{\phi\phi} \psi
    +\frac{1}{\rho}\partial_r\left(\rho\frac{f^2}{\Omega_{\rm BV}^2}\partial_r \psi\right)\right]+\frac{1}{r}\partial_{\phi} \psi\partial_y \Pi=0,
\end{equation}
where $\psi$ is the pressure perturbation and 
$f$ is the Coriolis frequency $2\Omega \cos \theta$. The angular components of the Lapalacian are noted $\Delta_{\theta\theta}A=\frac{1}{r^2}\frac{1}{\sin\theta}\partial_\theta\left(\sin\theta \partial_\theta A \right)$
and 
$\Delta_{\phi\phi}A=\frac{1}{r^2\sin^2\theta}\partial_\phi(\partial_\phi A)$.
$\Pi$ is the potential vorticity or vortensity of the background fluid, which is the vorticity divided by the density and multiplied by the potential temperature (a function of entropy and $Y_e$).
Here we change the coordinate of the equation from the original one (Eq.~(9) of \cite{Spruit84}) to be suited for our coordinate except for the last term since the notation of $\partial_y \Pi$ is widely used in the previous studies. The $y$-direction is  the $\theta$-direction in our coordinate system.

Considering perturbations defined by $\psi = \Psi(r,\theta) \exp\left(\imath m \phi - \imath \omega t\right)$, the equation leads to
\begin{equation}
\left[
\frac{\partial_\theta\left(\sin\theta \partial_\theta \psi \right)}{r^2\sin\theta}
+\frac{-m^2 \psi }{r^2\sin^2\theta}
+\cdots
\right]
+\frac{m\partial_y \Pi }{r  (m\Omega-\omega)} \psi =0.
\end{equation}
Here we omit the detailed expressions and defer the full discussion of its solutions to a future publication.
We regard this equation as a Schr{\"{o}}dinger-like equation,
$\frac{\partial^2\Psi}{\partial^2 x}=V_{\rm eff} \Psi$ in the $\theta$-direction.
The effective potential $V_{\rm eff}$ becomes
\begin{equation}
V_{\rm eff}=\frac{m}{r}\frac{\partial_y\Pi}{\omega-m\Omega}+\cdots.\label{eq:effectivepotential}
\end{equation}
To trap the Rossby wave near the corotation radius, $V_{\rm eff}$ should be negative and surrounded by positive values like a potential well \citep{Lovelace14}. The change of sign of $\partial_y\Pi$ is thus important to trap the wave.
Following Eq.~(18) of \cite{Spruit84} (their variable $\Lambda$ corresponds to our variable $90^{\circ}-\theta$), we obtain
\begin{align}
\partial_y \Pi &= 
\frac{2\Omega }{r}\sin\theta\left[
1+2\cos^2\theta\frac{r}{\rho}\partial_r \left(\rho\frac{\Omega^2}{\Omega_{\rm BV}^2}\alpha\right)
\right]
+\Delta_{\theta\theta}v_{\phi}
\label{eq:instability-criterion},
\end{align}
where $\alpha = -{{\rm d} \ln \Omega}/{{\rm d}\ln r}$.
The first term on the r.h.s. is the $\theta$-derivative of the Coriolis parameter, where the so-called $\beta$-approximation is used. 
To change the sign of the effective potential, the absolute value of the second term on the r.h.s. should be larger than the first term somewhere in $r$ and $\theta$, corresponding to a large gradient of $\rho \alpha\Omega^2/\Omega_{\rm BV}^2$.
Note that a similar criterion can be obtained without referring to an effective potential \citep{Spruit84}.

To summarize, a large gradient of density, angular frequency or \BV frequency
near the corotation radius is necessary to induce the instability.
Near the convective boundary, $|\Omega_{\rm BV}| < \Omega$ and a large gradient of $\rho \alpha\Omega^2/\Omega_{\rm BV}^2$ is expected.
Note that if the density and the \BV frequency are continuous, a strong shear (differential rotation) 
is necessary to trigger the instability.
This is similar to the analysis in the context of isentropic isolated neutron stars, where our Eq.~\eqref{eq:effectivepotential} and Eq.~\eqref{eq:instability-criterion} 
correspond to Eq.~(24) in \cite{Yoshida17}.

We confirm this scenario by evaluating $\partial_y\Pi$ of Eq.~\eqref{eq:instability-criterion} in the top left panel of Figure~\ref{fig:R1.0-3D-ss-80ms}.
This snapshot corresponds to 80 ms postbounce. $\partial_y\Pi$ is calculated from the azimuthal average of the hydrodynamic quantities.
The white line corresponds to the corotation radius of $\Omega/2\pi=125\,{\rm Hz}$.
Near $(R,Z)=(20\,{\rm km},25\,{\rm km})$ or $(15\,{\rm km},-25\,{\rm km})$, $\partial_y\Pi$ changes the sign, i.e., the value there is negative and at slightly lower-latitude, the value is positive.
The sign of the effective potential changes near the corotation radius.
Interestingly this change does not take place at the equator but at the mid-latitude $\theta\sim 60^{\circ}$ or $\sim 125^{\circ}$.
The radial positions of the regions are near the outer convective boundary. In the bottom left panel of Figure~\ref{fig:R1.0-3D-ss-80ms},
we show $\Omega_{\rm BV}$. The convectively unstable region is colored as red and yellow.
The regions with negative $\partial_y \Pi$ are extended from the outer convective boundary to the outer direction.

The emergence of the instability at mid-latitude is natural since the physical mechanism of the instability is similar to that of cyclones on Earth, which also requires sufficient Coriolis force and appears at mid-latitude.
The term inside the square brackets in Eq.~\eqref{eq:instability-criterion} has a form of $1-A\cos^2\theta$.
For a sufficiently large positive  value of $A$,
the term is negative at the pole and positive at the equator.
The change of sign takes place at mid-latitude.
The detailed properties in this snapshot are shown in middle and right panels of Figure~\ref{fig:R1.0-3D-ss-80ms}.
In the PNS, the negative gradient of $Y_e$ induces convective motions.
The top middle panel depicts $\delta Y_e=(Y_e-\langle Y_e\rangle_{\theta\phi})/\langle Y_e\rangle_{\theta\phi}$ at $r=25\,{\rm km}$, where the bracket means the polar and azimuthal angle average.
The panel shows 4 red (or white) regions at $\theta=120$-$150^{\circ}$.
These correspond to the (anti-)cyclones, which are correlated to an updraft, $v_r>0$.
Outside of them, vorticity spreads out due to the geostrophic balance, i.e., the pressure gradient force is cancelled by a Coriolis force and accelerated to rotate around the region.
Since the higher $Y_e$ region corresponds to a high-pressure (anti-cyclone), winds from the region are expected in the $\theta$-$\phi$ plane.
In the southern hemisphere, the anti-cyclone should rotate anti-clockwise (positive vorticity), like on Earth.
The top right panel shows the r-component of the vorticity and one can confirm the growth of the vorticity around the region.
In the southern hemisphere, the vorticity is basically negative (red), and the instability makes oppositely rotating positive vorticity (white) at $\theta=120$--$150^{\circ}$.
In the northern hemisphere, the background vorticity is basically positive (blue). However, one can also find
a slight $m=4$ component (white) at $\theta\sim 60^\circ$.
This is also made by anti-cyclone in the northern hemisphere, which is shown as the yellow region in the top middle panel.
In the panel, the snapshot at 82 ms is used to best show this $m=4$ structure clearly while the profiles of 80 ms are used in other panels.
In the vorticity panel (top right), a
belt ($m=0$) of negative vorticity (red)
is seen at $\theta\sim 45^{\circ}$.
This belt is also made by the geostrophic balance.
This region corresponds to the boundary of the higher $Y_e$ region, which is confirmed in the top middle panel.
A strong wind blows from the south to the north, deviated by the Coriolis force.

How does the instability at mid-latitude connect to the \ltw instability at the equator?
The bottom middle panel in Figure~\ref{fig:R1.0-3D-ss-80ms} shows the dispersion of angular velocity $\delta\Omega$ divided by average $\langle\Omega\rangle_\phi$ in the plane $R$-$Z$.
The dispersion quantify the deviation from axisymmetry, corresponding to the growth of $m\neq 0$ mode.
The dispersion is larger in the southern side ($Z<0$) since the vorticity in the southern hemisphere is stronger than in the northern hemisphere (see the top right panel).
The vorticity is mainly produced at mid-latitude and is transferred to the equator. In the convectively unstable zone, the stratification is weak and radial motions are allowed. Then the vortex becomes parallel to the rotation axis ($z$-axis) in the convective zone.
Finally in the $z=0$ plane, one can see the vortex.
In the bottom right panel, a mode $m=4$ appears clearly in the red belt at $r\sim20\,{\rm km}$.
This structure is produced by the $m=4$ mode at mid-latitude, discussed in the previous paragraph.

\begin{figure}
    \centering
	\includegraphics[width=0.8\columnwidth]{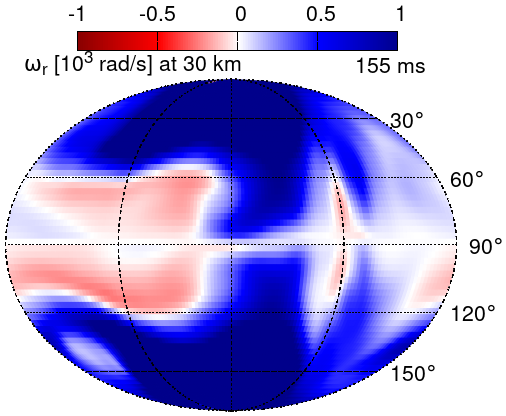}\\
	\includegraphics[width=0.8\columnwidth]{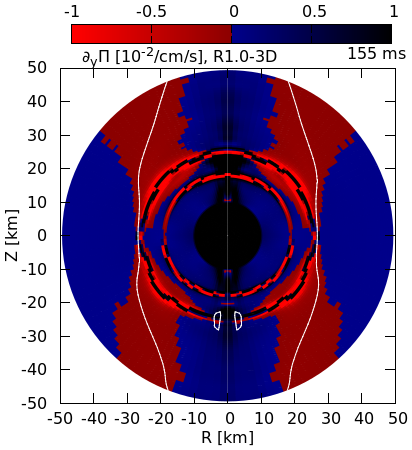}
    \caption{Top: $r$-component of the vorticity in $r=30\,{\rm km}$ plane.
    The Mollwide projection is used and the horizontal axis represents the longitude.
    The 5 tics are $0^\circ$, $90^\circ$, $180^\circ$, $270^\circ$ and $360^\circ$ from left to right. Different from the top right panel of Figure~\ref{fig:R1.0-3D-ss-80ms}, the sign of the variable in the southern hemisphere is changed to see the structure clearly, i.e., $-\omega_r$ is plotted in the southern hemisphere.
    Bottom: $\partial_y \Pi$ in $R$-$Z$ plane. The white line depicts the corotation radius of $\Omega/2\pi=110\,{\rm Hz}$. }
    \label{fig:R1.0-3D-ss-155ms}
\end{figure}

This clear $m=4$ structure soon disappears and
the distribution of $Y_e$ and vorticity becomes irregular.
Several modes appear to be mixed.
Finally at 155 ms, the mode $m=1$ becomes dominant.
In the top panel of Figure~\ref{fig:R1.0-3D-ss-155ms}, the vorticity is plotted using the Mollweide projection.
The horizontal axis represents the longitude, $\phi$, and the 5 tics correspond to $0^\circ$, $90^\circ$, $180^\circ$, $270^\circ$ and $360^\circ$ from left to right.
In the panel, the mode $m=1$ (red part) extends from $\phi=0$ to $180^\circ$ near the equator.
Similar results are reported in \cite{Takehiro02,Takehiro98} who
investigated rotational effect in thermal convection.
The results depend on the Rayleigh number and the Taylor number.
In their 2D simulation with high Rayleigh number, several convective plumes appear initially and  
they are merged to make a larger convective cell.
They argue that non-linear effects are important since 
the linear growth rate cannot explain the growth rate of the large convective cell.
Following the same argument, we speculate that our
$m=4$ mode at 80 ms grows following the linear theory and the dominance of the $m=1$ mode at 155 ms is the consequence of non-linear effects. Note that at 155 ms the $m=2$ mode coexists with the $m=1$ mode (see Figure~\ref{fig:R103Dt-f-d}).
That is also confirmed in the top panel of  Figure~\ref{fig:R1.0-3D-ss-155ms}. One can find a small red region at $\phi=270^\circ$, and with the big red region of $\phi=0{\rm -}180^\circ$, that makes $m=2$ mode.

What ingredient is the driver of the instability in the non-linear phase?
The bottom panel of Figure~\ref{fig:R1.0-3D-ss-155ms} showing $\partial_y \Pi$ reveals that 
the unstable region  defined by the red and blue boundary appears at low latitude.
This is consistent with the fact that the vortex appears at a low latitude in the top panel.
From these clues, we can infer that the driver of the instability in the non-linear phase could be similar to the linear phase, i,e., the Rossby wave is trapped in this region and generate the vorticity there.

\begin{figure}
	\includegraphics[width=0.9\columnwidth]{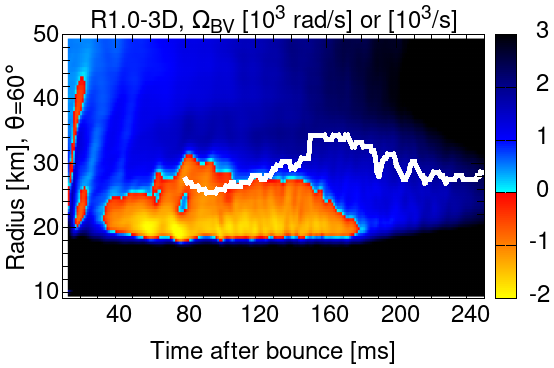}\\
	\includegraphics[width=0.9\columnwidth]{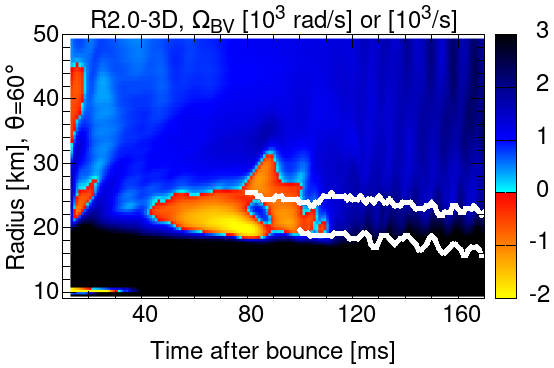}
    \caption{The space time diagram of \BV frequency. The vertical axis is radius at $\theta=60^\circ$. The white line represents the position of corotation radius. The top and bottom panel are for the slowly and rapidly rotating model, respectively. The white lines in the bottom panel are the positions of the corotation radii of the \ltw instability.}
    \label{fig:t-r-BV}
\end{figure}

We have shown the snapshots in the linear phase (80 ms) and the non-linear phase (155 ms).
Next we move to the time evolution of the corotation radius and $\Omega_{\rm BV}$. In the top panel of Figure~\ref{fig:t-r-BV},
the space-time diagram of the \BV frequency for the slowly rotating model is shown, which is calculated by the hydrodynamic variables averaged azimuthally. 
The white line is the position of the corotation radius.
Unlike Figures \ref{fig:R1.0-3Dpropagation} and \ref{fig:R2.0-3Dpropagation}, we now consider the quantities at $\theta=60^\circ$ where the instability is triggered.
    
In the linear phase (80-150 ms), the position of the corotation radius is almost at the outer convective boundary. In the non-linear phase (after 150 ms), the azimuthally averaged \BV frequency might not be a good indicator of the convection due to non-linear or non-axisymmetric effects. The boundary between the red and the blue regions is no longer correlated to the white line. However, this is common in convective turbulence. The initial negative $Y_e$ or entropy gradient tends to be relaxed after the onset of strong convective mixing. Though the negative gradient is smoothened, we may expect that the convection continues in the region. The bluish region at 20-30km after 160 ms has a smaller \BV frequency compared to the other regions and may correspond to the convective region. In this sense, the corotation radius is always located near the outer convective boundary in this model.

\begin{figure*}
	\includegraphics[width=0.3\linewidth]{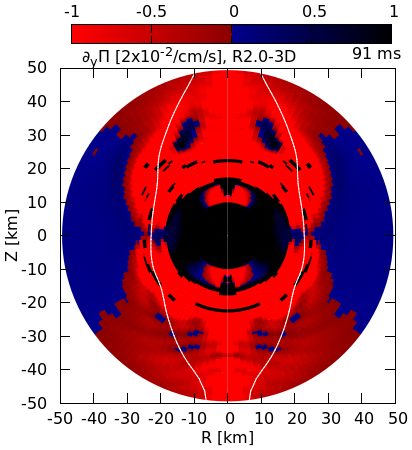}
	\includegraphics[width=0.3\linewidth]{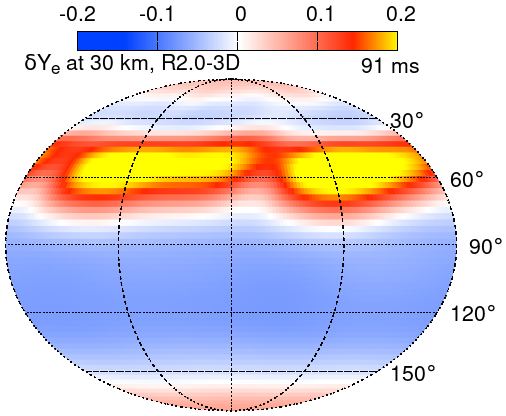}
	\includegraphics[width=0.3\linewidth]{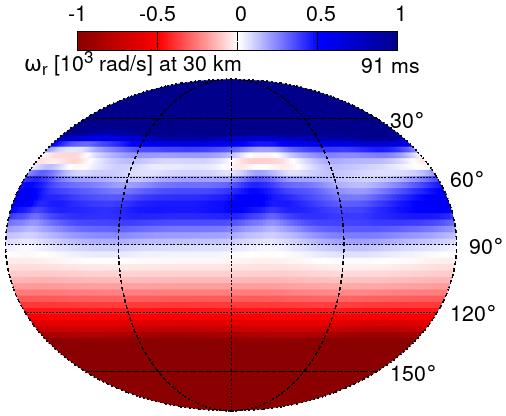}\\
	\includegraphics[width=0.3\linewidth]{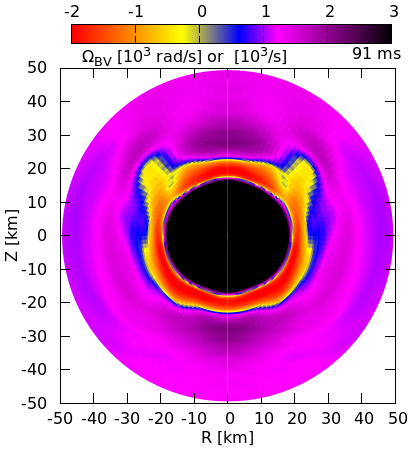}
	\includegraphics[width=0.3\linewidth]{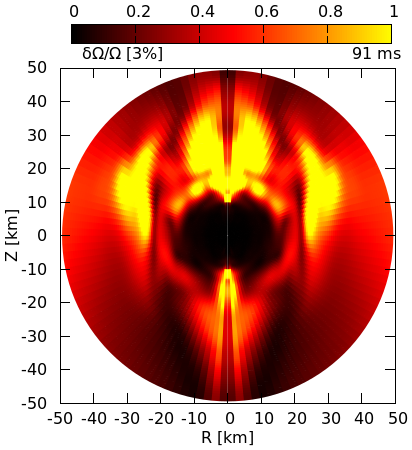}
	\includegraphics[width=0.3\linewidth]{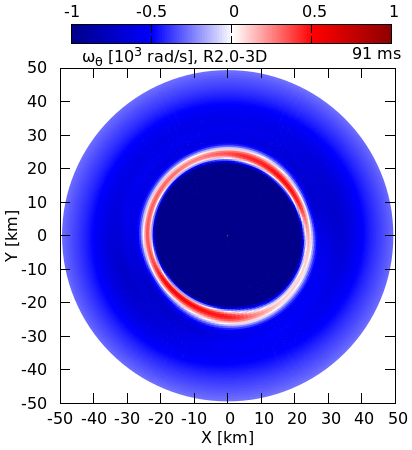}
    \caption{Same to Figure~\ref{fig:R1.0-3D-ss-80ms} but for R2.0-3D model. The snapshots of $91\,{\rm ms}$ are taken. }
    \label{fig:R2.0-3D-ss91ms}
\end{figure*}

We also visit the rapidly rotating model where some features are similar to that of the slowly rotating model.
Figure~\ref{fig:R2.0-3D-ss91ms} shows several snapshots of R2.0-3D model at the 91ms, the early epoch of \ltw instability.
In the top left panel, $\partial_y\Pi$ is evaluated. At $(R,Z)=(20\,{\rm km},30\,{\rm km})$, the bright red region appears, and that is next to the blue region at $(R,Z)=(30\,{\rm km},25\,{\rm km})$
In these regions, the sign of $\partial_y\Pi$ changes.
The region is near the corotation radius of $\Omega/2\pi=200\,{\rm Hz}$ (white line).
This meets the criterion of the instability. The bright red region begins from the convective layer and extends to the outer region beyond the outer convective boundary. The bottom left panel shows the \BV frequency. One can confirm the relation between $\partial_y\Pi$ and $\Omega_{\rm BV}$. The convective region is colored as red and yellow there.

In this phase, the mode $m=2$ appears. The top middle panel of Figure~\ref{fig:R2.0-3D-ss91ms} shows
the Mollweide projection of $\delta Y_e/\langle Y_e\rangle_{\theta\phi}$. $m=2$ pattern is found at $\theta\sim 60^\circ$.
The two big yellow regions are anti-cyclones (high-pressures) whose center is near $\phi\sim90^\circ$ and $\sim 270^\circ$ (2nd and 4th tics from the left, respectively).
From the geostrophic balance,  vorticity is raised around the anti-cyclones.
The top right panel shows the r-component of the vorticity.
The white belt at $\theta=\sim 60^\circ$ is made by the anti-cyclones.
Interestingly, strong shear is found between the two anti-cyclones near $\phi=0^\circ$ and $180^\circ$.

The strong vortex ($\theta\sim 60^\circ$) at the surface of the sphere ($r=30\,{\rm km}$) is connected to a vortex in the $z=0$ plane.
The bottom middle panel of Figure~\ref{fig:R2.0-3D-ss91ms} shows the deviation from axi-symmetry of the angular frequency $\delta \Omega/\langle \Omega \rangle_{\phi}$. The deviation is strong in the northern part ($Z>0$) and correlated with the unstable region shown in the top left panel. The strong dispersion continues vertically (along $z$-axis) and reaches the equatorial plane. 
The bottom right panel shows the $\theta$-component of the vorticity.
The red ring at $R=\sqrt{x^2+y^2}\sim20\,{\rm km}$ is related to the belt at $\theta\sim 60^\circ$ in the top right panel.
The $m=2$ structure of the ring is strong at $\phi\sim90^\circ$ and $\sim 270^\circ$.
This coincides with the position of the anti-cyclones in the top middle panel.

\begin{table*}
\caption{}
\begin{tabular}{llllll}
\hline
  References     &  Site        & Geometry        & Baroclinicity & Solberg-Høiland  & \BV \\
  \hline
  \cite{Lovelace14} & disks & equatorial plane & barotropic & stable & unstable\\
  \cite{Lyra19}     & disks & equatorial plane & baroclinic & stable & unstable\\
  \cite{Yoshida17}  & stars & equatorial plane & barotropic & stable & stable\\
  \cite{Spruit84}   & stars & sphere           & baroclinic & stable or marginal & stable or marginal\\
           \hline
\label{tab:differencesoftheworks}
\end{tabular}
\end{table*}

\begin{figure}
    \centering
	\includegraphics[width=0.8\columnwidth]{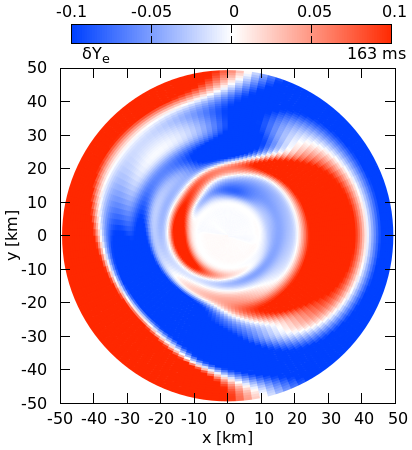}\\
	\includegraphics[width=0.8\columnwidth]{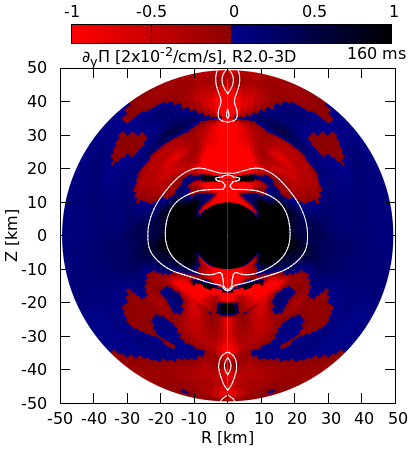}
    \caption{Top: $\delta Y_e$ at $z=0$ plane.
    Bottom: $\partial_y \Pi$ in $R$-$Z$ plane. The white line depicts the corotation radius of $\Omega/2\pi=110\,{\rm Hz}$. }
    \label{fig:R2.0-3D-ss-160ms}
\end{figure}

Soon the non-linear phase comes ($t>100\,{\rm ms}$), and the $m=1$ mode becomes dominant.
Interestingly, two $m=1$ modes appear in this phase (see Figure~\ref{fig:R20t-f-pat} and Eqs. \eqref{eq:f_lowinR2.0}-\eqref{eq:f_hghinR2.0}).
The shape of the modes are clearly depicted in the top panel of Figure~\ref{fig:R2.0-3D-ss-160ms}, which shows $\delta Y_e$ at 163ms in the $z=0$ plane. Here $\delta Y_e\equiv (Y_e-\langle Y_e\rangle_\phi)/\langle Y_e \rangle_\phi$.  
In the outer region $R=20$-$40\,{\rm km}$, the large excess (red part) is found at $x>0$ and $|y|<30\,{\rm km}$. 
On the other hand, in the inner region $R<20\,{\rm km}$, the excess is located at  $x<0$ and $|y|<20\,{\rm km}$. 
Both modes are $m=1$, and naturally have different frequencies since the inner region rotates more rapidly than the outer region. The stability criterion using $\partial_y \Pi$ is perhaps helpful to understand the origin of two $m=1$ modes.
The bottom panel of Figure~\ref{fig:R2.0-3D-ss-160ms} shows the value of $\partial_y \Pi$ at 160 ms.
The two white lines denote the two corotation radii. The outer and inner lines correspond to $200$ and $300$ Hz, respectively. The frequencies are close to the frequency of the \ltw instability . Along the outer white line, the sign of $\partial_y \Pi$ changes at $(R,Z)=(20\,{\rm km},10\,{\rm km})$.
This region might be responsible for the origin of the outer $m=1$ mode.
The position is near the outer convective boundary and similar to the slowly rotating model. 
On the other hand, along the inner white line, the sign of $\partial_y \Pi$ changes at  $(R,Z)=(10\,{\rm km},15\,{\rm km})$.
This region may be responsible for the instability in the inner region.
We do not have any good explanation for the structure of $\partial_y \Pi$ in the bottom panel.
More studies are necessary to pin down the origin of the high-frequency mode in the inner region.
For example, pressure gradient in the $z=0$ plane can be important \citep[see Section 4.4.2 of ][for the case of protoplanetary disk]{Lyra19}.
A caveat to keep in mind is that we assume spherical transport for $r<10\,{\rm km}$ and do not include the transport to the $\theta$ and $\phi$ directions, so that the numerical time step does not become too small. This treatment may affect the hydrodynamic structure here.

The low frequency mode of the rapidly rotating model appears near the outer convective boundary. 
It is similar to that of the slowly rotating model. 
The bottom panel in Figure~\ref{fig:t-r-BV} shows the \BV frequency in the space time diagram.
Since the instability begins at mid-latitude, we choose the radius at $\theta=60^\circ$ for the vertical axis.
Before 100 ms, the convective region is clearly seen as red and yellow. 
The outer white line is the corotation radius of the low frequency mode
and clearly corresponds to the outer convective boundary. As discussed before, in the non-linear phase, the relation between the convective layer and corotation radius becomes less obvious since the non-linear effect mixes the matter in the convective layer and makes $|\Omega_{\rm BV}|$ smaller. However, we point out that the position of the corotation radius in non-linear phase is still near the convective layer. As discussed in Figure~\ref{fig:R2.0-3D-ss-160ms}, the high frequency mode is triggered by the complex structure in the polar region. 
Though the instability seems to happen near the inner convective boundary in the bottom panel of Figure~\ref{fig:t-r-BV},
we do not know whether this is a general feature or not.

To summarize this section, we have pointed out that the \ltw in PNS is triggered by a Rossby wave instability at mid-latitude
where the meridional derivative of the potential vorticity, $\partial_y\Pi$ changes its sign.
Though a higher $m$ mode appears in the linear phase,
a strong $m=1$ mode in the equator is seen in the non-linear phase.
The evolution of the vorticity is governed by a geostrophic balance, i.e., the balance between Coriolis force and the pressure gradient force. The geostrophic balance naturally predicts vorticity that surround high or low pressure region.
It is similar to the cyclones or anti-cyclones on Earth.

The Rossby wave instability in our study has unique features though we have stressed the similarity to other works in the beginning of this section. 
Table~\ref{tab:differencesoftheworks} summarizes the similarity and differences among the previous works. One essential feature of \cite{Spruit84} is that they consider a sphere in a star. Generally speaking, stratification is strong in stars. The $r-$component of the vorticity, $\omega_r$, is important since it is parallel to the direction of the gravitational force \citep[see][for the detail of this quasi-geostrophic approximation]{Pedlosky82}. In this case, the planetary $\beta$-effect should be considered, i.e., the relative vorticity changes as a function of $\theta$ since the strength of the Coriolis force changes. At the equator in disks or stars, whose stratification in the $r$-direction is weak compared to rotational effects, the $\theta$-component of the vorticity, $\omega_\theta$, is important. That is affected by the topographic $\beta$-effect, i.e., the typical height of the vortex tube changes as a function of radius. Though the governing equations of these systems are similar, they should not be confused. The component of the Coriolis force also explains the difference in the geometry. The full form of the Coriolis force is written as $(f_r,f_\theta,f_\phi)= 2m\Omega(0,-v_\phi \cos\theta,v_\theta \cos\theta-v_r \sin\theta)$ in spherical coordinates.  In the planetary $\beta$-effect, $v_r=0$ is assumed since the strong stratification suppresses the radial motion. Then $(f_\theta,f_\phi)= 2m\Omega\cos\theta(-v_\phi,v_\theta)$ remains as the popular form of the Coriolis force. This force vanishes at the equator, $\theta=90^\circ$. On the other hand, at the equator of less stratified stars or disks,  $v_\theta=0$ is assumed. Still $f_\phi= -2m\Omega\sin\theta v_r$ remains. Even if $\theta=90^\circ$, this term does not vanishes. In this case, the centrifugal force may affect the motion of the fluid.
 
Another important aspect is the effect of the equation of state.  \cite{Lovelace14} and \cite{Yoshida17} consider the barotropic situation where pressure is written as a function of density. On the other hand, \cite{Lyra19} and \cite{Spruit84} consider the baroclinic situation where the pressure is determined by density and temperature or $Y_e$.  In this case, 
the surface of constant density and constant pressure do not coincide in general.  This baroclinic effect is the origin of the second term of r.h.s in Eq.~\eqref{eq:instability-criterion} \citep{Spruit84}, which is the most important term to induce the \ltw instability in PNSs.

As already noticed,
the \ltw instability in PNS is quite unique and different from \ltw in disks and cold isolated NSs. In cold NSs, the differential rotation mostly drives the \ltw instability \citep[e.g.][]{Shibata03a}.
On the other hand, the existence of the convective layer seems more important in PNSs. Similar instability is perhaps known in the context of protoplanetary disks \citep{Lyra19}, but the geometry is different from the stellar sphere and we should not confuse the topographic $\beta$-effect and planetary $\beta$-effect. In disks, the unstable \BV frequency is smaller than the stable epicyclic frequency. The system is stable in the sense of the Solberg-Høiland criterion. In PNSs, the epicyclic frequency cannot stabilize the unstable \BV frequency. That is another difference from the instability in the disks.

\subsubsection{Effect of spiral SASI}\label{sec:sasispiral}

In another context, 
\cite{Kazeroni17} pointed out the connection between the \ltw instability and spiral SASI.
They have systematically investigated the impact of the strength of rotation on the 
growth rate of the non-asymmetric instability \citep[see also][]{Yamasaki08,Blondin17}, in which an idealized setting was taken where the PNS interior is excised and replaced by a fixed boundary.
  The growth rate of the $m=2$ mode slightly exceeds that of the $m=1$ mode, as the initial specific angular momentum is increased \citep{Yamasaki08}. 
In our case, the specific angular momentum behind the shock, $L$, is $0.6$-$0.8\times10^{16}\,{\rm cm^2/s}$ in model R1.0-3D with the $m=2$ mode deformation.
This is consistent with the Figure 2 in the linear analysis of \cite{Kazeroni17}, where 
the growth rate of the mode $m$=2 is higher than $m$=1 when $L \gtrsim 0.5\times 10^{16}\,{\rm cm^2/s}$. In what follows, we discuss the interplay between the spiral SASI and the \ltw instability by taking model R1.0-3D as a reference.

The evolution of the physical quantity $A$  perturbed by the spiral SASI can be expressed as
\begin{align}
    \delta A \propto \sin\left(2\pi f_{\rm s} t +\phi \right)\exp(\omega_{\rm s} t),\label{eq:sasipert}
\end{align}
where $f_{\rm s}$ is the angular frequency and $\omega_{\rm s}$ is the growth rate.
The subscript "s" refers to the spiral SASI.

\begin{figure}
	\includegraphics[width=\columnwidth]{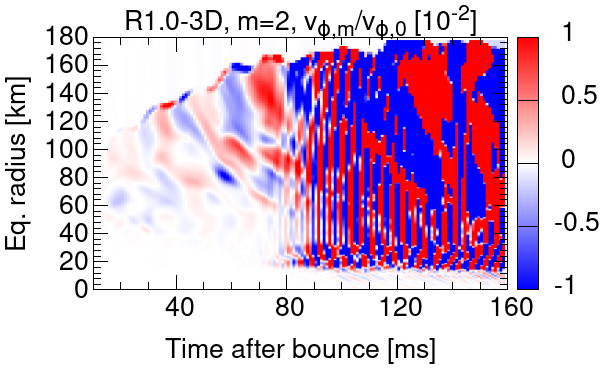}\\
	\includegraphics[width=\columnwidth]{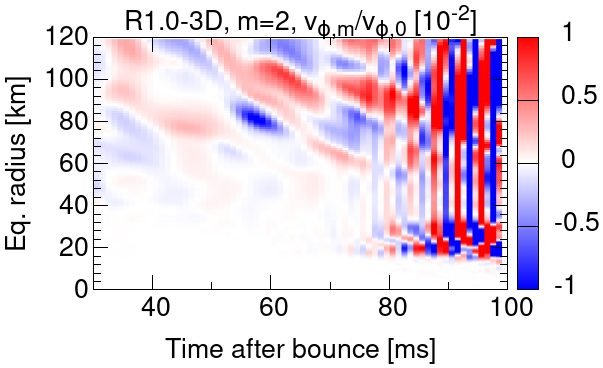}\\
	\includegraphics[width=\columnwidth]{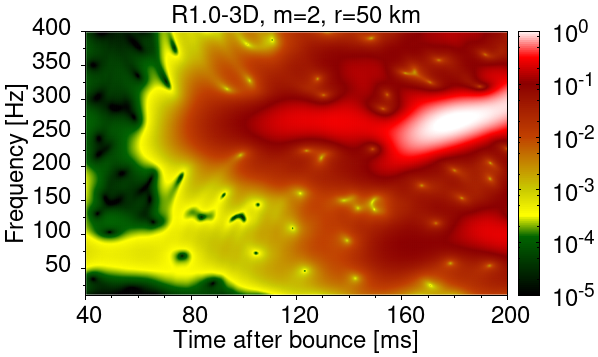}
	 \caption{Top: Space-time diagram of $v_{\phi,2}/v_{\phi,0}$ for model R1.0-3D.
	          See Eq.~\eqref{eq:vp-hamonics} for the definition.
             Note that the color scale is normalized by $10^{-2}$ in the panel.
	          Middle: Same as the top panel but zooming up $30\,{\rm ms}<t<100\,{\rm ms}$ and $0\,{\rm km}<r<120\,{\rm km}$.
           Bottom: The spectrogram of $v_{\phi,2}/v_{\phi,0}$ at the equatorial radius 50\,km.}
    \label{fig:R103Dt-r-Vpm2}
\end{figure}

From Figure~\ref{fig:R103Dt-r-Vpm2}, we estimate $f_{\rm s}$.
The top panel shows
the space-time diagram of harmonic decomposition of the azimuthal velocity (e.g. Eq.~(\ref{eq:rho-hamonics})):
\begin{align}
v_{\phi,m}(t,r) =& \int{\rm d}\phi\,v_\phi(t,r,\theta=\pi/2,\phi) \cos(m\phi).\label{eq:vp-hamonics}
\end{align}
The middle panel is the same as the top panel, but  focuses on the early evolution of $t_{\rm pb} \sim$ 30--100\,ms for the more central region at the equatorial radius of 0-120\,km.
In the bottom panel,  we show the spectrogram of $v_{\phi,2}$. The feature after $t_{\rm pb} \sim $80\,ms is similar to that of the density (see Figure~\ref{fig:R103Dt-f-d}, bottom panel).
From the spectrogram (bottom panel), we can estimate the oscillation frequency of the pattern, 
i.e., $f_{\rm s}\sim$  75\,Hz, which is shown as a yellowish horizontal excess at $t_{\rm pb} \lesssim$ 80 ms in the panel.

\begin{figure}
	\includegraphics[width=\columnwidth]{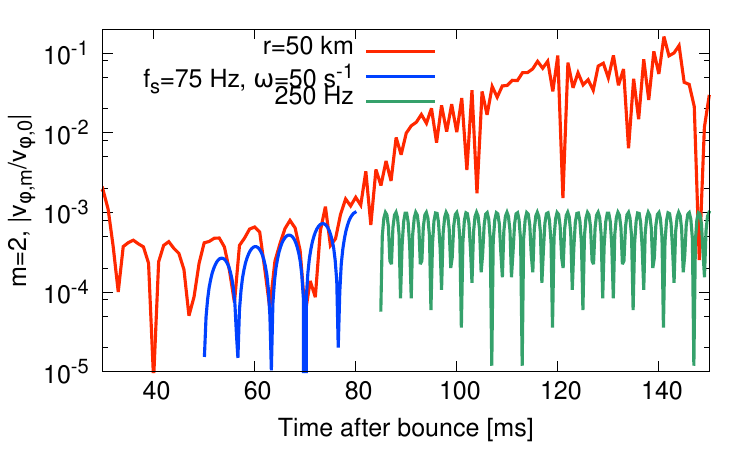}
	 \caption{
	 The time evolution of $|v_{\phi,2}/v_{\phi,0}|$ at 50\,km for model R1.0-3D is shown by the red curve. The blue curve is 
	  Eq.~\eqref{eq:sasipert} with $f_{\rm s}=75\,{\rm Hz}$ and $\omega_{\rm s}=50\,{\rm s^{-1}}$. The green curve is a sin-curve of 250\,Hz.
	  }
    \label{fig:R103Dr-tadv}
\end{figure}

Next, we roughly estimate the growth rate.
In the top panel of Figure~\ref{fig:R103Dr-tadv}, we plot $v_{\phi.2}$ at 50 km as a function of the postbounce time (red line).
At $t_{\rm pb} \lesssim$ 80 ms, the red curve is well fitted by Eq.~\eqref{eq:sasipert} with $f_{\rm s}=75\,{\rm Hz}$ and $\omega_{\rm s}=50\,{\rm s^{-1}}$ (blue curve).

After $t_{\rm pb} \sim$ 80ms, the stripe pattern in the top panel of Figure~\ref{fig:R103Dt-r-Vpm2} is significantly different from $t_{\rm pb} \lesssim$ 80ms.
As shown before,
this oscillation is due to the \ltw instability and not due to the spiral SASI.
First, the pattern frequency of the \ltw instability is 250\,Hz (for the $m$=2 mode, bottom panel of Figure 6),
and that of the SASI is 75\,Hz (see the bottom panel of Figure~\ref{fig:R103Dt-r-Vpm2}) ).
Second, the region of the oscillation is also different:
the development of the \ltw instability begins at $\sim$ 20\,km (see the middle panel of Figure~\ref{fig:R103Dt-r-Vpm2}).
In the case of the SASI, the perturbed region is significantly higher up.

Before closing this section, we shall refer to a caveat in the above analysis.
In our models, the \ltw instability develops into the non-linear regime before the spiral SASI does. However, the dominance between the \ltw instability and the spiral SASI should depend primarily on the employed progenitor models, and also on the assumed initial angular momentum. If the initial rotation rate is smaller, the spiral SASI can be the dominant mode \citep{Walk18}, where the \ltw instability is of minor importance.
Further investigation is necessary to clarify how the mechanism of the dominant instability is determined for a wide variety of progenitors with rotation, and also ultimately with magnetic fields.

\begin{figure*}
	\includegraphics[width=\columnwidth]{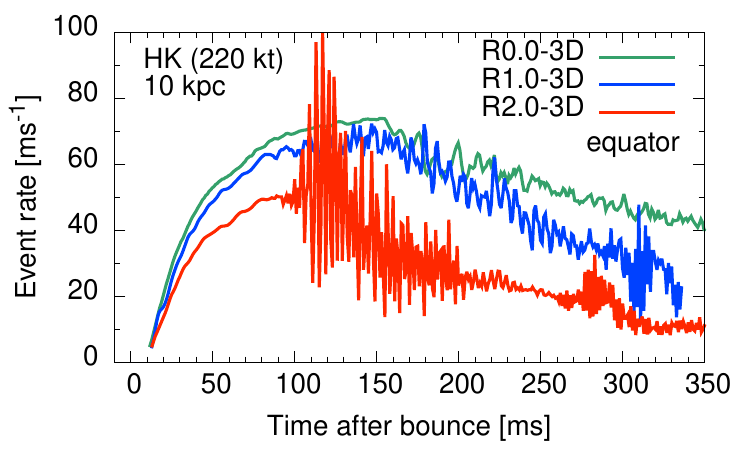}
	\includegraphics[width=\columnwidth]{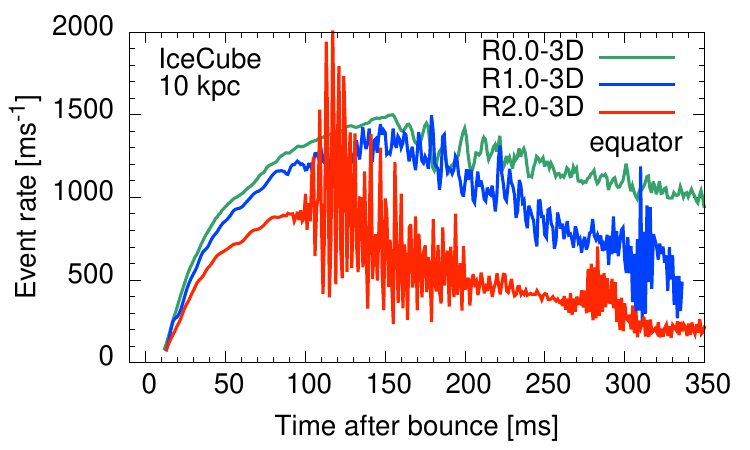}\\
	\includegraphics[width=\columnwidth]{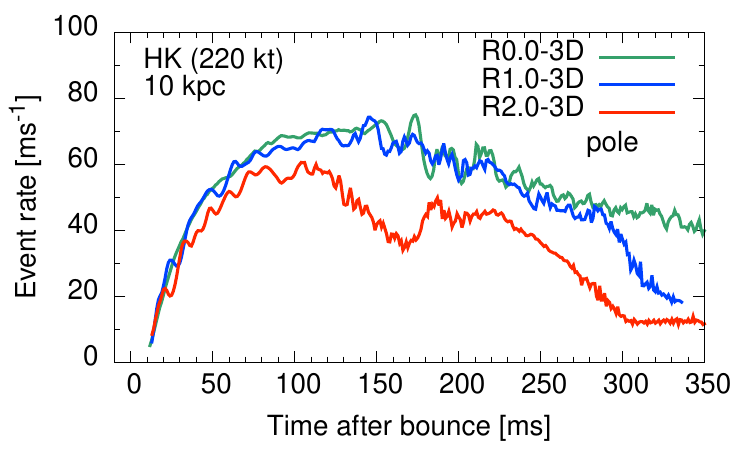}
	\includegraphics[width=\columnwidth]{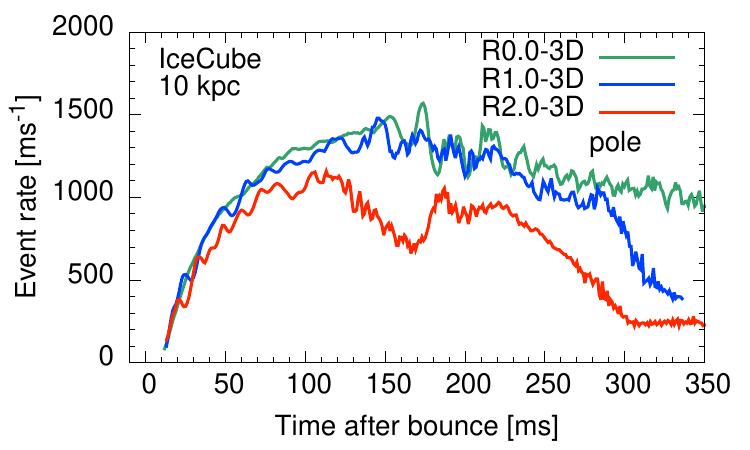}
    \caption{Expected detection rates of $\bar{\nu}_e$ at Hyper-Kamiokande (HK) and IceCube at a source distance of $10$ kpc in the left and right panels, respectively. The rates for the observer along the equatorial (polar) direction are shown in the top (bottom) panels. The red, blue, and green lines are for models R2.0-3D, R1.0-3D, and R0.0-3D, respectively.  }\label{fig:t-neutrinoevent}
\end{figure*}

\subsection{Neutrino and Gravitational-wave Signatures}\label{sec:GW_nu}
We analyse the neutrino and gravitational-wave (GW) signatures from our models in Sections \ref{nu} and \ref{gw} respectively. Focusing on the impact of the \ltw instability, we also discuss the detectability of the multi-messenger signals.

\begin{figure}
	\includegraphics[width=\columnwidth]{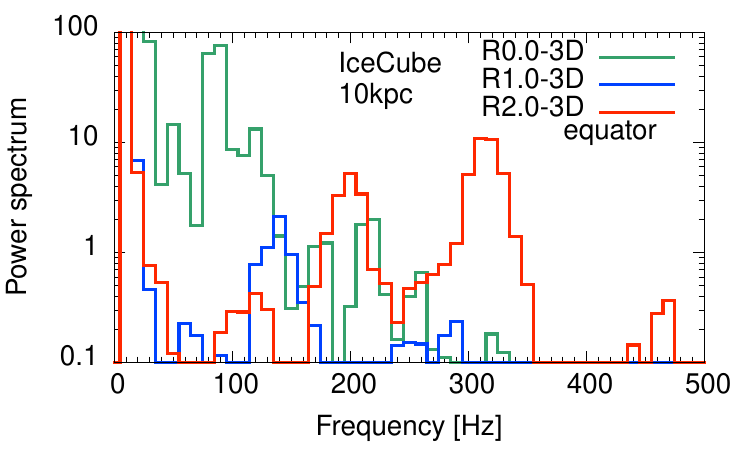}
    \caption{Fourier power spectrum of the IceCube event rates. The spectrum is normalized by the noise. The time-window, $\tau$, is $100{\rm ms}$ and the number of bin, $N_{\rm bin}$ is $100$. Note the difference of the time-intervals chosen for each model (see text). The red, blue, and green lines are used for models R2.0-3D, R1.0-3D and R0.0-3D, respectively.  }    \label{fig:f-neutrinoevent}
\end{figure}
\subsubsection{Neutrino signatures}\label{nu}

Two water \v{C}erenkov neutrino detectors, 
Hyper-Kamiokande \citep[HK,][]{HyperK2018} and
IceCube \citep[IC,][]{ICECUBE2011,ICECUBE2012} are considered as in \cite{Takiwaki18}.
In the detectors, the main detection channel is inverse-beta decay
 (IBD)  of anti-electron neutrino ($\bar{\nu}_e$).
 The observed event rate at HK is calculated as follows,
\begin{equation}
 R_{\rm HK} = N_{\rm p}
  \int_{E_{\rm th}} {\rm d} E_e
   \frac{{\rm d} F_{\bar{\nu}_e}}{{\rm d} E_\nu}
   \sigma\left(E_\nu\right)
   \frac{{\rm d} E_\nu}{{\rm d} E_e},
\end{equation}
where $N_{\rm p}=1.48\times 10^{34}$ is the number of protons for
 the tank of HK whose fiducial volume is designed as 220 kton.
The neutrino number flux of $\bar{\nu}_e$ ($F_{\bar{\nu}_e}$) at a 
source distance $D$ is 
estimated as $F_{\bar{\nu}_e} = \frac{\mathcal{L}_{\Omega}}{4\pi D^2}$, where $\mathcal{L}_{\Omega}$ 
denotes the viewing-angle dependent neutrino (number) luminosity \citep{Tamborra14prd}.
$D$ is set as $10$ kpc.
We assume a Gamma-distribution for the neutrino spectrum \citep{Tamborra14prd,Tamborra12}.
Note that in the previous paper of \cite{Takiwaki18}, we used 440 kton as the fiducial volme following the old plan
and employed a Fermi-Dirac distribution to reconstruct the spectrum. Other parameters 
including the threshold energy $E_{\rm th}=7~{\rm MeV}$ are the same as those in \citet{Takiwaki18}. We estimate the event rate in IceCube following \cite{Lund10}.
It should be noted here that our neutrino prediction is based on the ray-by-ray approximation \citep{Takiwaki16}, which will overestimate the viewing-angle variations. More accurate angle-dependent neutrino transport (e.g., \citet{Sumiyoshi12,Harada19}) is needed for a more quantitative prediction.

The time evolution of the event rate 
in HK (left panels) and IceCube (right panels) is shown in Figure~\ref{fig:t-neutrinoevent}
 for all the 3D models computed in this study. The top and bottom panels correspond to an observer in the direction parallel to the equator or along the pole, respectively.
 Comparing the top with bottom panels, the time variability in both the HK and IceCube events in models R2.0-3D (red line) and R1.0-3D (blue line) is strongest for the equatorial observer. In the rapidly rotating model R2.0-3D (red line), the strong signal modulation is seen from $t_{\rm pb} \sim $100 ms to 200 ms (top panels). In fact, the \ltw instability is fully developed for the time period (see Figure~\ref{fig:R20t-f-pat}).
The amplitude of the signal modulation is as large as its non-oscillating component ($\sim 40$ and $\sim 800$ ${\rm ms}^{-1}$ for HK and IceCube, respectively). 

The neutrino signal modulation produced by the growth of the \ltw instability is highly dependent on the viewing angle (compare the top panels with the bottom panels of Figure~\ref{fig:t-neutrinoevent}). This light-house effect was previously identified in \citet{Takiwaki18}: the spinning of strong neutrino emission regions around
the rotational axis leads to quasi-periodic modulation in the neutrino signal. 

Our slowly rotating model R1.0-3D also shows the neutrino signal modulation due to the \ltw instability.
From Figure~\ref{fig:R103Dt-f-d}, the non-linear phase sets in from $t_{\rm pb} \sim$ 120 ms.
The neutrino event rate also shows the variability from that time
(blue lines in the top panels of Figure~\ref{fig:t-neutrinoevent}).
The amplitude of the variability is not as large as in the rapidly rotating model.

A small neutrino variability is also seen in our non-rotating model, due to SASI rather than the \ltw instability. 
The green line in the top and bottom panel shows an oscillatory feature from 
$t_{\rm pb} \sim$ 150 ms. This SASI-induced neutrino modulation is consistent with those observed in \cite{Tamborra14prd,Walk18,Walk20}.

To obtain a spectral feature of the signal modulation in IceCube, we perform a 
Fourier analysis following \cite{Lund10}.
The important parameters are the signal duration $\tau$, the sampling time $\Delta$ and the number bin $N_{\rm bin}=\tau/\Delta$. In this work, we employ
$\tau = 100 {\rm ms}$, $\Delta=1 {\rm ms}$, $N_{\rm bin}=100$.
The time window is different in each model: $t_{\rm pb} =$ 100--200 ms for model R2.0-3D, 150--250 ms for model R1.0-3D, 150--250 ms for model R0.0-3D. The time window is chosen to focus on the epoch when the growth of the strong \ltw instability is observed (see the top panels of Figure~\ref{fig:t-neutrinoevent}). 
The resultant power spectrum is shown in Figure~\ref{fig:f-neutrinoevent}. The vertical axis is normalized by the noise level.

The top panel of Figure~\ref{fig:f-neutrinoevent} shows that model R2.0-3D has two peaks of $\sim 200\,{\rm Hz}$ and  $\sim 320\,{\rm Hz}$ in the neutrino spectrum. Interestingly, they correspond to the two frequencies of the \ltw instability, $f_{{\rm mode},1,{\rm low}}$ and $f_{{\rm mode},1,{\rm high}}$, in Eqs. \eqref{eq:f_lowinR2.0} and \eqref{eq:f_hghinR2.0}.
The amplitude of the two peaks (for a 10 kpc source) is larger than the noise level, which corresponds to unity in the $y$ axis of the figure. From the blue line in the figure, we can see that our slowly rotating model R1.0-3D has a peak at $\sim$ 130\,Hz which is roughly consistent with $f_{{\rm mode},1}$ in Eq.~\eqref{eq:f1inR1.0}.
The peak amplitude for this model slightly exceeds the detection limit of unity.
 Our non-rotating model (R0.0-3D, red line) has a strong peak at $\sim$ 90\,Hz, which corresponds to the frequency of SASI.  These features are qualitatively similar to those in our previous work \citep{Takiwaki18}, whereas the peak frequencies of the signal modulation is higher in this work. For example, the spectrum of model R2.0-3D in our previous study showed a single peak around $\sim 120$\,Hz. The difference should mainly come from our update in the treatment of gravity with the effective GR potential, while the purely Newtonian gravity was used in \citet{Takiwaki18}.
Since the evolution of the PNS is also affected by neutrino cooling, our update in the treatment 
of $\nu_X$ from the leakage scheme to the self-consistent IDSA scheme in this work should also 
contribute to the quantitative difference.

\begin{figure}
	\includegraphics[width=\columnwidth]{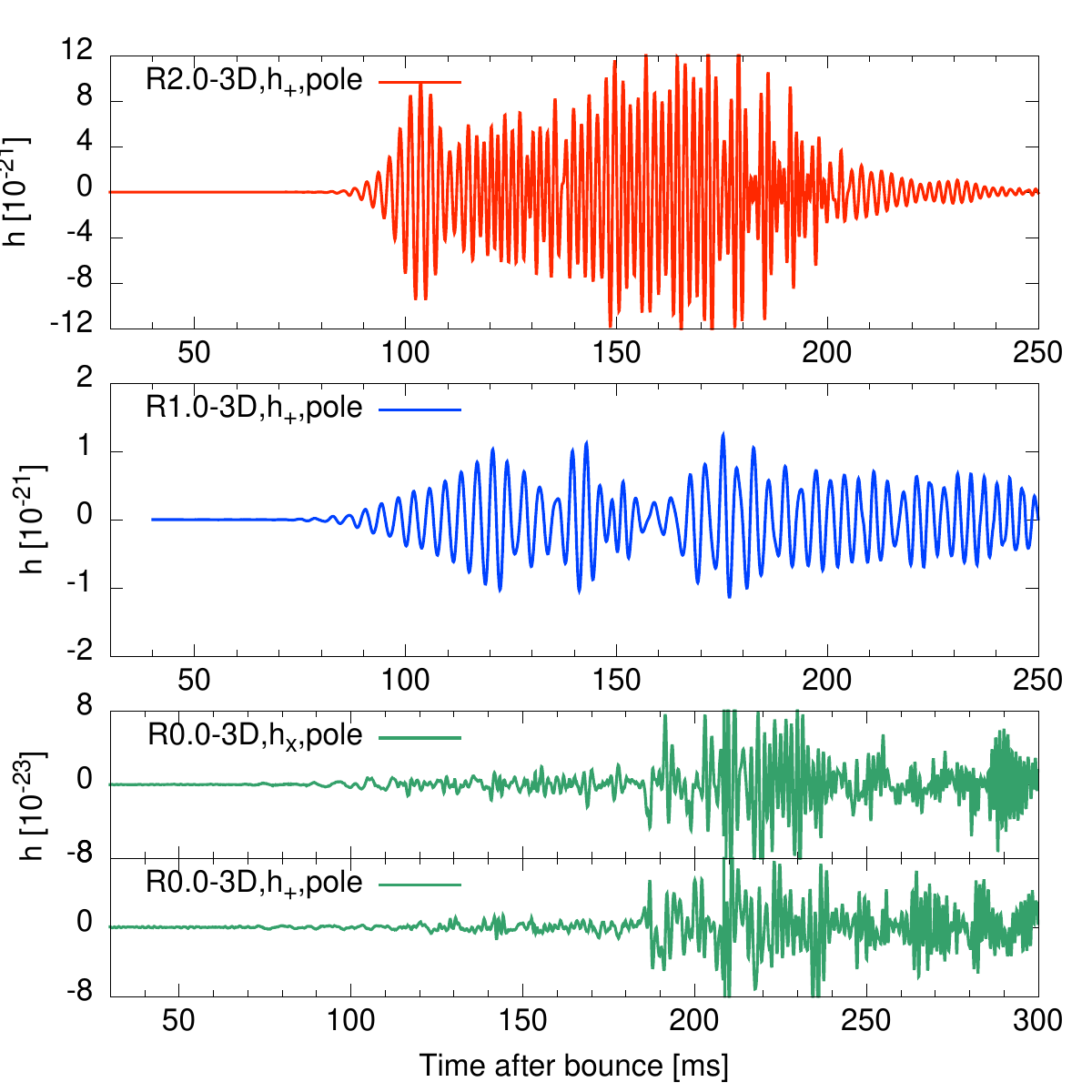}
    \caption{The gravitational waveforms of our models. The red, blue and green lines correspond to models R2.0-3D, R1.0-3D, and R0.0-3D, respectively. 
    The observer is located at a distance of 10 kpc along the $z$-axis (i.e. the spin axis for the rotating models). Note that the $y$-scale in the bottom panel is 100 times smaller than that of the rotating models. 
    The data is available at \url{https://doi.org/10.5281/zenodo.5489428}.
    }
    \label{fig:t-h_gw}
\end{figure}

\subsubsection{Gravitational-wave signatures}\label{gw}
The gravitational-wave (GW) signals are also expected to reflect directly the instabilities growing in the supernova core (see \citet{ernazar2020,Kotake13} for reviews).
In this section, we present the waveform analysis focusing on the connection to the \ltw instability and also discuss the detectability of the GW signals.

We extract the gravitational waveform using the quadrupole formula.
We first evaluate the first moment of the momentum density, $\dot{I}_{ij}^{\rm TT}$
using Eq.~(37) of \cite{Finn90},
\begin{equation}
  \dot{I}_{ij}^{\rm TT}=
  2 \int {\rm d}V
  \rho \left( \frac{v_i x_j + v_j x_i}{2} - \frac{\delta_{ij} v_k x_k }{3} \right)
\end{equation}
every $\Delta t=0.1\,{\rm ms}$ in the simulations, where the spatial indices $i$ and $j$ run from $1$ to $3$ ($x$ to $z$).
Then we numerically differentiate $\dot{I}_{ij}^{\rm TT}$ to obtain the waveform i.e.,
\begin{align}
h_{ij}
=\frac{1}{D}\frac{G}{c^4}\frac{{\rm d}^2 }{{\rm d}^2 t} I_{ij}^{\rm TT}
=\frac{1}{D}\frac{G}{c^4}\frac{{\rm d} }{{\rm d}t} \dot{I}_{ij}^{\rm TT}
=\frac{1}{D}\frac{G}{c^4}\frac{\left. \dot{I}_{ij}^{\rm TT}\right|_{n+1} - \left. \dot{I}_{ij}^{\rm TT}\right|_{n}}{\Delta t},
\label{quad}
\end{align}
where $D$ is the distance of the source and $\left.\right|_{n}$ indicates the $n$-th timestep estimated at a time interval of $\Delta t$,
$G$ is the gravitational constant, and $c$ is the speed of light.
If the supernova source is observed along the rotation axis \citep{EMuller12},
the two polarized amplitudes are related as
\begin{align}
h_{+} = h_{xx} - h_{yy},\ h_{\times} =  2 h_{xy}.
\end{align}
The gravitational waveform of our models are shown in Figure~\ref{fig:t-h_gw}.
Note for the rotating models (R1.0-3D and R2.0-3D), we plot only the $+$-mode since the waveform of the $\times$-mode is similar to the $+$-mode except for the trivial phase shift. 

As shown in the top panel,
the GW amplitude of model R2.0-3D becomes bigger after the onset of the \ltw instability, at $t_{\rm pb} \sim 100\,{\rm ms}$. The wave amplitude for this model changes more strongly with time compared to the slower rotating model (middle panel). From the middle panel, we can see the strong GW emission of the \ltw instability, $100\,{\rm ms}<t_{\rm pb}$. As previously identified in \citet{andresen19}, model R0.0-3D (green line in the bottom panel) shows that the waveform is less sensitive to the 
viewing angle, because the growth of SASI as well as convection have no preferential direction in the non-rotating model.
In what follows, we explore how the GW signatures can be related to the growth of the \ltw instability for our rotating models. 

\begin{figure}
	\includegraphics[width=\columnwidth]{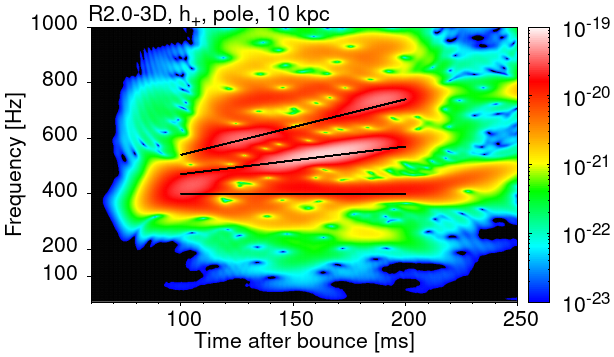}\\
	\includegraphics[width=\columnwidth]{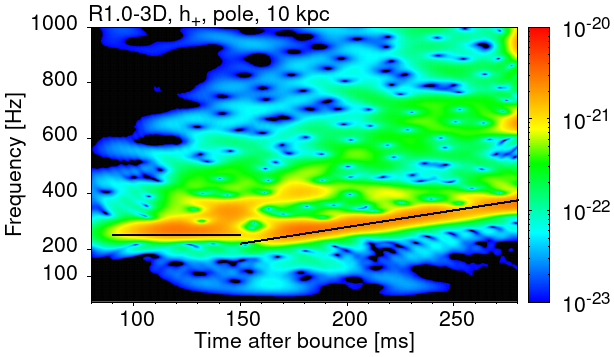}\\
    \caption{Spectrograms of the characteristic GW amplitude, $h_{\rm char}$ for models R2.0-3D (top panel) and R1.0-3D (bottom panel), respectively.
     The observer is located at a distance of 10 kpc along the rotational axis.
The thick black lines are drawn to characterize the excess in the spectrogram.
In the top panel, the three black lines correspond to the characteristic frequencies of $2f_{{\rm mode},1,{\rm high}}$, 
$f_{{\rm mode},1,{\rm high}}+f_{{\rm mode},1,{\rm low}}$, and
$2f_{{\rm mode},1,{\rm low}}$, from high to low, which are defined in Eqs. \eqref{eq:f_lowinR2.0} and \eqref{eq:f_hghinR2.0}. In the bottom panel, 
 the black line corresponds to
 $f_{{\rm mode},2}$ in Eq.~\eqref{eq:18} and 
 $2f_{{\rm mode},1}$ in Eq.~\eqref{eq:f1inR1.0}.}
    \label{fig:t-f-h_gw}
\end{figure}

\begin{figure}
	\includegraphics[width=\columnwidth]{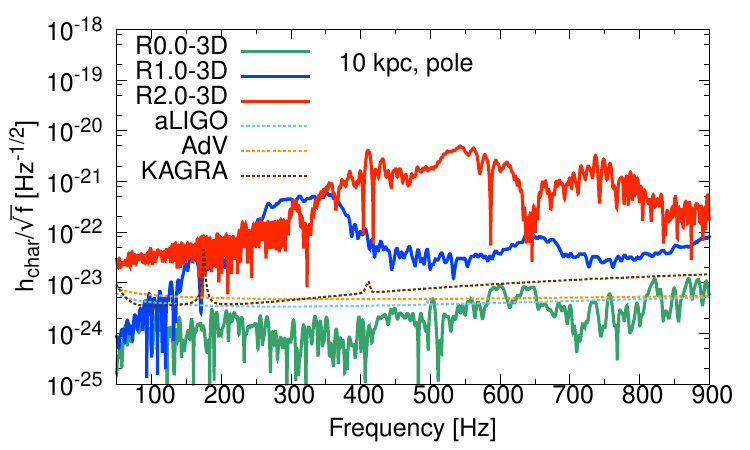}
    \caption{The characteristic strain of the GW signals from our 3D models, relative
     to the sensitivity curves of advanced LIGO (aLIGO), advanced VIRGO (AdV), and KAGRA
      (the data is extracted from the following URL,
     \url{https://gwdoc.icrr.u-tokyo.ac.jp/cgi-bin/DocDB/ShowDocument?docid=9537}).
     An observer is assumed to be located at a distance of 10 kpc along the rotation axis of the source. }
    \label{fig:f-h_gw}
\end{figure}

To show the spectral evolution of the GW signals,
we plot in Figure~\ref{fig:t-f-h_gw} the spectrogram of the characteristic amplitude $h_{\rm char}$ \citep[e.g.][]{Murphy09},
\begin{align}
    h_{\rm char}(t_{\rm pb},f) &= 
    \sqrt{\frac{2G}{\pi^2 c^3 D^2} \frac{{\rm d} E_{\rm GW}}{{\rm d }f}},\\
    \frac{ {\rm d} E_{\rm GW} }{ {\rm d} f}(t_{\rm pb},f) &= \frac{3G}{5 c^2} \left(2\pi f\right)^2 
    \left| \tilde{S}\right|^2,\\
    \tilde{S}(t_{\rm pb},f) &= \frac{1}{2} 
    \int_{t_{\rm pb}-\Delta T}^{t_{\rm pb}+\Delta T} {\rm d}t^{\prime} \frac{c^4}{G} D h(t^\prime) W(t^\prime-t_{\rm pb})\exp\left(-2\pi {\rm i}  ft^\prime \right),\label{eq:shortftgw}\\ 
W(t^\prime-t_{\rm pb}) &=\left[1+\cos\left(\frac{\pi\left(t^\prime-t_{\rm pb}\right)}{2\Delta T}\right)\right],
\end{align}
where we take the sampling time $\Delta T$ as 20 ms.
 
 The bottom panel of Figure~\ref{fig:t-f-h_gw} shows that the characteristic GW frequency (guided by the thick black line) increases with time. This is in accordance with the increase of the mode frequency due to the \ltw instability 
 (see Figure~\ref{fig:R103Dt-f-d}). In the bottom panel, the thick black line corresponds to 
 $f_{{\rm mode},2}$ (Eq.~\eqref{eq:18}) and 
 $2f_{{\rm mode},1}$ (Eq.~\eqref{eq:f1inR1.0}), which shows a good agreement with the excess in the GW spectrogram (red region). As already pointed out in \citet{Takiwaki18}, the GW frequency produced by the spiral ($m=1$) mode of the \ltw instability  is two times higher than the $m=1$ modulation frequency due to the frequency doubling, inherent to the quadrupole nature of GW emission.
  The following relation, $2f_{{\rm mode},1}=f_{{\rm mode},2}$, should be also kept in mind. Not surprisingly, the GW frequency is the same as the $m=2$ modulation frequency, provided that $m=2$ is the dominant deformation mode of the PNS (see also \citet{Shibagaki20}). 
  
  These features of the GW frequency can also be interpreted in an analytical manner as discussed in Appendix~\ref{sec:gwtoymodel}. From Eq.~~\eqref{eq:gwtoyslow}, one can see that the \ltw instability preferentially leads to the GW emission with the frequency $2\Omega$ (where $\Omega$ is the angular velocity of the mode) for model R1.0-3D.
In the rapidly rotating model R2.0-3D, the GW
frequencies can be expressed as a linear combination of the PNS deformation modes. As already shown in the top panel of Figure~\ref{fig:t-f-h_gw}, the three black lines correspond to $2f_{{\rm mode},1,{\rm high}}$, 
$f_{{\rm mode},1,{\rm high}}+f_{{\rm mode},1,{\rm low}}$, and
$2f_{{\rm mode},1,{\rm low}}$ from top to bottom. 
The three lines deduced from Eqs. \eqref{eq:f_lowinR2.0} and \eqref{eq:f_hghinR2.0} almost agree with the characteristic frequencies (red region). Our analytical estimate gives a good agreement of the three lines  $2\Omega_1,\Omega_1+\Omega_2,2\Omega_2$ (Eq.~\eqref{eq:gwtoyrapid}) by the superposition of the modes. The origin of the power excess at a higher frequency than $2 \Omega_1$ would require a more dedicated analysis of the non-linear mode coupling. We leave this for future study.

Finally we discuss the detectability of the GW signals in our models.
Following the literature \citep{Murphy09,Takiwaki18,Shibagaki20}, we compare the characteristic GW strain $h_{\rm char}$ with the detector sensitivity curves (Figure~\ref{fig:f-h_gw}). Note in this plot that the window-function $W$ is set to unity and we  integrate the amplitude over the simulation-time in Eq.~\eqref{eq:shortftgw}.

The slowly rotating model R1.0-3D (the blue line in Figure~\ref{fig:f-h_gw}) has a peak around 300 Hz which comes from the mode of the \ltw instability and is expected in the bottom panel of Figure~\ref{fig:t-f-h_gw}. The crude estimate of the signal to noise ratio ($S/N$) is given by the ratio of the signal and the sensitivity. In this case, $S/N \sim 100$ should be detectable for a source at 10 kpc.
In the case of the rapidly rotating model, R2.0-3D (the red line), the broad-band (400 - 800 Hz) spectrum is primarily produced by the sum of several modes as seen in the top panel of Figure~\ref{fig:t-f-h_gw}. The $S/N$ is also $\sim 100$, which is within the detection limits of advanced detectors for a source at 10 kpc. To claim detection, $S/N$ should be greater than $\sim 10$ \citep[see][and the references therein]{Hayama15}, in this sense, the source within 100\,kpc would be detectable.

\section{Summary and Discussions}\label{sec4}

We have presented a detailed analysis aiming to obtain deeper insights into 
 the mechanism of the \ltw instability in the context of rapidly rotating CCSNe. To this end, we have performed 3D core-collapse simulations of a $27 M_{\odot}$ star including several important updates in the GR correction to gravity, the multi-energy treatment of heavy-lepton neutrinos via the IDSA scheme, and the use of a more realistic EOS by \citet{togashi}, improving on the assumptions of our previous study \citep{Takiwaki18}. In this work, we have computed three models, where the initial angular momentum is parametrically added to the progenitor core.
 The shock revival was obtained in the rotating models as reported in \cite{Takiwaki16}.
 
 We first focused on the evolution of the pattern frequency and corotation radius, which has been considered important to understand the 
 nature of the \ltw instability in the isolated cold neutron stars. In the analysis, we have found that the corotation radius appears near the convective layer in the PNS.
 This property led us to propose a novel triggering mechanism of the \ltw instability in the CCSN environment.
 Following the linear analysis of \cite{Spruit84}, the non-axisymmetric Rossby wave grows near the convective layer where the \BV frequency is sufficiently small.
 Since the Rossby wave is ruled by the Coriolis force, the instability first appears at mid latitude.
 With non-linear effects, large scale azimuthal modes ($m=1$ or $2$) finally survive and become synchronized in both hemispheres, leading to the large scale spiral arms extending across the equator, as often observed in numerical simulations of the \ltw instability.
 We point out that this mechanism might work even in progenitors with relatively slow rotation. We observed the growth of the \ltw instability at $T/|W|\sim 0.02$ at bounce (in our model R1.0-3D).
 
 We have also investigated how the growth of the \ltw instability impacts the neutrino and  GW     signatures and discussed its detectability in advanced LIGO, advanced Virgo, and KAGRA.
 Strong time variability was obtained in the neutrino signal, whose amplitude is strongest for observers in the equatorial direction. The GW frequency is basically twice as high as the PNS rotation rate. If there are several hydrodynamics modes that concurrently account for the PNS deformation, the GW frequencies were shown to 
  be related to the linear combination of these modes.

Although we show several shreds of evidence for the new scenario,
we consider only two models. Computing more  3D models (possibly with a more idealized setup) would be helpful to understand further the nature of the \ltw instability in the PNS.
By applying the linear perturbation analysis proposed in the cold NSs 
 \citep[e.g.][]{Passamonti15,Yoshida17},  one could, in principle, identify the nature of the trapped wave. However, this is not an easy task because of the presence of the 
 shock, convection, and the SASI in the CCSN environment.  Note 
 that previous simulations using a simplified setup did not include the effect of convection 
  \citep[e.g.][]{Shibata03a,Ou06,Kazeroni17}.
The effect of convection is indispensable for producing the instabilities observed in our study.
Whether the \ltw instability or the spiral-SASI is dominant might depend on the initial angular momentum. Further study, such as the parametric modeling proposed by \citet{Kazeroni17}, is needed to answer this question.

So far, we have not considered the effects of magnetic fields, which can become dynamically important especially in rapidly rotating progenitors.
Strong magnetic fields may suppress the \ltw instability \citep{Fu11,Muhlberger14}.
The impact of the magnetorotational instability (MRI) on the explosion mechanism is also a long-standing issue \citep{Akiyama03, Masada15,Sawai16,Guilet15b,Moesta15,Rembiasz16,Reboul20}.
The MRI should certainly affect the angular momentum profile and affect the \ltw instability \citep{Bugli18}.
The impact of the magnetic fields on convection is currently gaining more attention \citep{JMatsumoto20,Muller20b,Raynaud20,Masada20}.
3D magnetohydrodynamics simulations (e.g., \citet{Obergaulinger20,Kuroda20}) should be performed in order to obtain a more complete description of the \ltw instability.

\section*{Acknowledgements}
We thank Y. Masada and N. Yokoi for useful and stimulating discussions.
Numerical computations were
carried out on Cray XC50 at the Center for Computational Astrophysics,
National Astronomical Observatory of Japan.
This work was supported by Research Institute of
Stellar Explosive Phenomena at Fukuoka University and the associated project (No. 207002), and also by JSPS KAKENHI
Grant Number
(JP17H01130, 
JP17K14306, 
JP17H06364, 
JP18H01212, 
and JP21H01088). 
This research was also supported by MEXT as “Program for Promoting 
researches on the Supercomputer Fugaku” (Toward a unified view of 
the universe: from large scale structures to planets) and JICFuS.

\section*{Data Availability}

The data underlying this article will be shared on reasonable request to the corresponding author.




\bibliographystyle{mnras}
\bibliography{mybib} 



\appendix

\section{Simplified Analytical Formulae for the gravitational emission from the \ltw instability}\label{sec:gwtoymodel}
 From the standard quadrupole formula of the GW emission \citep{EMuller12}, one can estimate the GW amplitude emitted toward the rotational ($z$) axis as
 \begin{align}
    h &= \frac{2G}{c^4 D}\int{\rm d}V \rho\left(v_y v_y -v_x v_x
 -y\partial_y \Phi
+x\partial_x \Phi\right).
\label{quada}
 \end{align}
 In what follows, we neglect the contribution from the derivatives of the gravitational potential (the third and fourth terms in the R.H.S of the above equation) because the rotational energy is dominant for our rapidly rotating model. Note in the non-rotation cases, these terms should not be neglected because their contribution are almost the same as the first two terms related to the kinetic energy.
 We assume a symmetry in the $z$-direction in the cylindrical geometry and impose the (sinusoidal) perturbations produced by the \ltw instability
  as,  
  \begin{align}
  \rho &= \rho_0\left(1+\delta \rho(\phi)\right),\nonumber\\
  v_y &= v_{\rm rot}\left(1+\delta v(\phi))\right)\cos\phi,\nonumber\\ 
  v_x &= v_{\rm rot}\left(1+\delta v(\phi)\right)\sin\phi,
 \end{align}
  where unperturbed values of $\rho_0, v_{\rm rot}$ are assumed to be constant.
 Putting these quantities into Eq.~(\ref{quada}), the GW amplitude is then written as, 
 \begin{align}
 h\propto \int{\rm d}\phi
\left(1+\delta \rho \right)
\left(1+\delta v \right)^2
\cos 2\phi.
\label{perturb1}
 \end{align}
In our slowly rotating model (R1.0-3D), the density and velocity perturbation induced by the \ltw instability can be expressed as,
  \begin{align}
 \delta \rho = \epsilon_{\rho} \cos (m(\Omega t -\phi),\nonumber\\
\delta v = \epsilon_{v} \cos (m(\Omega t -\phi)),
\label{perturb2}
\end{align}
 where $\Omega$ is the angular velocity and $\epsilon_{\rho}, \epsilon_v$ denotes the degree of the asymmetry of the density and velocity fields.
  Inserting Eqns. (\ref{perturb2}) to Eq.~(\ref{perturb1}), 
  the GW amplitudes resulting from (the mixture of) the $m=1$ and $m=2$ modes (see the middle panel of Figure~2) are expressed as, 
  \begin{align}
  h\propto
  \begin{cases}
    \frac{\pi}{2} \epsilon_v \left(2\epsilon_\rho+\epsilon_v\right) \cos\left(2 \Omega t \right) & m=1 \\
    \frac{\pi}{4} 
    \left[8\epsilon_v+
      \epsilon_\rho \left(4+3\epsilon_v^2\right)\right] \cos\left(2 \Omega t \right) & m=2,  
  \end{cases}\label{eq:gwtoyslow}
\end{align}
respectively. For the $m=1$ mode, the GW frequency is twice as high as the rotational frequency, whereas the GW frequency of the $m=2$ mode equals to the rotational frequency.
From Figures \ref{fig:R103Dt-f-d} and \ref{fig:R103Dt-r-Vpm2},
we can roughly estimate the degree of asymmetry in the non-linear phase ($t_{\rm pb} > 150\ {\rm ms}$): $\epsilon_\rho\sim 0.1$ and $\epsilon_v\sim 1$.
Substituting these values, we anticipate that the GW amplitude from the $m=1$ mode is as large as that of the $m=2$ mode. More importantly, we point out that the GW frequency is characterized by $2\Omega$ even if the $m=1$ and $m=2$ perturbation modes coexist.
 
 A similar analysis can be made for our rapidly rotating model (R2.0-3D).
 Since two $m=1$ modes coexist with different frequencies (see Figure~\ref{fig:R203Dt-r-d}),
 we take the following perturbations:
 \begin{align}
 \delta \rho &= \epsilon_{\rho,1} \cos (\Omega_1 t -\phi)
+\epsilon_{\rho,2} \cos (\Omega_2 t -\phi), \nonumber\\
\delta v &= \epsilon_{v,1} \cos (\Omega_1 t -\phi)
+\epsilon_{v,2} \cos (\Omega_2 t -\phi).
\end{align}
 The resultant GW amplitude is 
 \begin{align}
 h\propto \frac{\pi}{2}
  & \left[ \epsilon_{v,1} \left( 2\epsilon_{\rho,1} + \epsilon_{v,1}\right) \cos\left(2 \Omega_1 t \right)  \right.\nonumber\\
 +& \left. \epsilon_{v,2} \left(2\epsilon_{\rho,2} + \epsilon_{v,2}\right) \cos\left(2 \Omega_2 t \right)   \right.\nonumber\\
 +& \left.
2\left( \epsilon_{\rho,1}\epsilon_{v,2} + \epsilon_{\rho,2}\epsilon_{v,1}+\epsilon_{v,1}\epsilon_{v,2}\right) \cos\left((\Omega_1+\Omega_2) t \right)
\right].\label{eq:gwtoyrapid}
\end{align}
Due to the superposition of the two modes, the three frequencies of the GW emission appear at the frequencies $2\Omega_1$,$2\Omega_2$, and $\Omega_1+\Omega_2$.  This can explain the power excess in Figure~
\ref{fig:t-f-h_gw} (top panel), corresponding to the three black thick lines. 

\bsp	
\label{lastpage}
\end{document}